\documentclass[prb,twocolumn,floatfix]{revtex4}

\usepackage{amsfonts}
\usepackage{amsmath}
\usepackage{verbatim}
\usepackage[dvips]{graphicx}
\usepackage{subfigure}

\def\dd{\textrm{d}}

\def\sst{\scriptstyle}

\begin{document}

\title{Interaction proximity effect at the interface between a superconductor and a topological insulator quantum well}
\author{Predrag Nikoli\'c$^{1,2}$ and Zlatko Te\v{s}anovi\'{c}$^{2}$}
\affiliation{$^1$School of Physics, Astronomy and Computational Sciences,\\George Mason University, Fairfax, VA 22030, USA}
\affiliation{$^2$Institute for Quantum Matter at Johns Hopkins University, Baltimore, MD 21218, USA}
\date{\today}


\begin{abstract}

A material whose electrons are correlated can affect electron dynamics across the interface with another material. Such a ``proximity effect'' can have several manifestations, from order parameter leakage to generated effective interactions. The resulting combination of induced electron correlations and their intrinsic dynamics at the surface of the affected material can give rise to qualitatively new quantum states. For example, the leaking of a superconducting order parameter into certain Rashba spin-orbit-coupled materials has been recently identified as a path to creating ``topological superconductors'' that can host Majorana particles of use in quantum computing. Here we analyze the other aspects of the superconducting proximity effect. The proximity-induced interactions are a promising path to incompressible quantum liquids with non-Abelian fractional quasiparticles in topological insulator quantum wells, which could also find applications in topological quantum computing. We discuss the operational and design principles of a heterostructure device that could realize such states. We apply field-theoretical methods to characterize the properties of induced interactions via the electron-phonon coupling and Cooper pair tunneling across the interface. We argue that bound-state Cooper pairs can be stabilized by the interaction proximity effect inside a topological insulator quantum well at experimentally observable energy scales. The condensation of spinful triplet pairs, enabled by the Rashba spin-orbit coupling and tunable by gate voltage, would lead to novel superconducting states and fractional topological insulators.

\end{abstract}

\maketitle

\section{Introduction}

Interfaces between materials with radically different electronic ground states are increasingly promising platforms for engineering novel quantum states. The quest for elusive Majorana particles has been driving the latest wave of interest in such interfaces. Majorana quasiparticles were first predicted to exist as special Andreev bound states at the boundaries or inside vortex cores of spinless weakly-coupled $p_x + i p_y$ superconductors\cite{Read2000}, and similarly at the terminal points of quantum wires\cite{Kitaev2000}. However, it seems more practical to seek analogous Majorana particles in superconducting states with time-reversal (TR) symmetry that can be artificially created in systems with a strong spin-orbit coupling \cite{Fu2008, Sato2009}. The solid-state materials of this kind are not natural superconductors, but pairing can be induced via the ``proximity effect'', by placing them in contact with a conventional superconductor\cite{Fu2008}. Numerous theoretical proposals of proximity effect devices utilizing surfaces of the bulk topological insulators\cite{Fu2008, Stanescu2010a, Sau2010c, Linder2010, Sato2010} (TI), films\cite{Sau2010, Sau2010a} and quantum wires\cite{Lutchyn2010, Oreg2010, Stanescu2011, Cook2011, Alicea2011, Potter2012, Lin2012, Fidkowski2012, Kells2012, Stanescu2012a, Alicea2012} have been made and accompanied by microscopic model calculations\cite{Stanescu2010a, Valls2010, Lababidi2011, Rol2011, Sau2012, Michelsen2012}. The experimental search for Majorana particles based on those ideas\cite{Mourik2012, Deng2012, Rokhinson2012, Das2012} and the exploration of proximity effects\cite{Kasumov1996, Koren2011, Qu2011, Sacepe2011, Yang2011, Zhang2011, Veldhorst2012, Wang2012, Wang2012a, Williams2012, Yang2012} are also gaining momentum.

Consider an interface between a correlated material C whose ground state spontaneously breaks some symmetry (such as a superconductor) and an uncorrelated material UC whose ground state is naturally disordered (such as a metal or band-insulator). The term ``proximity effect'' is usually used to describe the leakage of the order parameter across the interface from C to UC. The material C presents an explicit symmetry-breaking perturbation to the electron dynamics inside UC, so that some electron correlations of the kind found in C will in most cases appear near the interface in UC (if its susceptibility to the perturbation is finite). In this manner correlations can be imposed on electrons in UC where they do not naturally occur. This phenomenon is behind all proposals to obtain Majorana particles in TIs and similar systems.

In this paper we qualitatively analyze other ways in which a correlated material C can induce correlations among electrons in a naturally uncorrelated material UC. Our interest is the ``interaction proximity effect'', or the renormalization of electron interactions in UC by their coupling to the degrees of freedom in C. We will specifically consider an interface with a conventional superconductor (SC) and explore two sources of effective interactions in the two-dimensional electron system (2DES) formed near the interface in the material UC: the direct coupling of the 2DES electrons to the SC's phonons, and Cooper pair tunneling from the SC to the 2DES. The second mechanism incorporates the usual ``proximity effect''. We will describe the proximity-induced interactions in 2DES using interaction potentials and effective actions, and discuss their ability to generate new states of matter with spontaneously broken symmetries or topological order.

\begin{figure}
\includegraphics[width=2in]{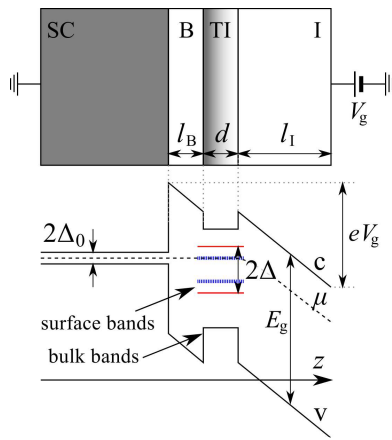}
\caption{\label{SCTI}(color online) The heterostructure device that embeds a topological insulator quantum well (TI) between a conventional superconductor (SC) and a conventional insulator (I). The chemical potential $\mu$ in the quantum well can be controlled by applying a gate voltage $V_{\textrm{g}}$, and more efficiently so if a tunneling barrier (B) is inserted at the SC-TI interface. A schematic band-structure of this device is also shown. Bulk band edges are drawn as long solid black lines, while the two-dimensional band edges of the quantum well are drawn as short solid red lines separated by $2\Delta$. These two-dimensional bands are created by the hybridization of surface states that would be gapless and topologically protected in a bulk three-dimensional TI. Their surface hybridization bandgap is smaller than the bandgap of a bulk TI. The blue dotted lines are the two-dimensional sub-gap Cooper pair bands created by the interaction proximity effect.}
\end{figure}

The current motivation to study the proximity-induced interactions comes from the possibility of obtaining fractional topological insulator states in TI quantum wells\cite{Nikolic2011a, Nikolic2012a}, which we briefly review here. Consider a heterostructure in the Fig.\ref{SCTI} which consists of a TI quantum well sandwiched between a bulk superconductor (SC) and a conventional insulator (I). The TI-SC interface can be bridged by a tunneling barrier (B), but this may not be necessary. This setup allows one to control the chemical potential inside the insulating TI quantum well by the gate voltage $V_{\textrm{g}}$ and hence drive quantum phase transitions at low temperatures. We will consider quantum wells made from ``strong'' TI materials such as Bi$_2$Se$_3$ or Bi$_2$Te$_3$. Electrons inside the quantum well have the spin $\sigma^z$ and a two-state ``orbital'' degree of freedom $\tau^z$. If we neglect their coupling to the rest of the heterostructure, then the basic aspects of their two-dimensional dynamics are captured by the single-particle Hamiltonian written in terms of the spin $\sigma^{x,y,z}$ and orbital $\tau^{x,y,z}$ Pauli matrices:
\begin{equation}\label{Bernevig}
H_0 = \frac{v}{2}\,\hat{{\bf z}}(\boldsymbol{\sigma}\times{\bf p})\tau^{z}+\Delta\tau^{x}-\mu \ ,
\end{equation}
where $v \sim 10^5 \textrm{ m/s}$ is the strength of the Rashba spin-orbit coupling\cite{Hsieh2009}, and $\Delta \sim 10 - 100 \textrm{ meV}$ is a bandgap\cite{Zhang2010, Zhang2009c, Kong2010, Hong2010, Liu2011a, Cho2011a} formed by the hybridization between the helical Dirac quasiparticles from the opposite surfaces $\tau^z = \pm 1$ of the TI quantum well. Note that this model also captures the Dirac ``helical'' spectrum of topologically protected surface states in thick TIs, where $\Delta = 0$. By continuity, this ensures the model's validity at least for sufficiently thick TI quantum wells (experimentally, a few quintuple layers).

The proximity-induced interactions among the TI's electrons of any spin can create two-dimensional Cooper pairs in several channels. Particularly interesting are the inter-orbital spin-triplet pairs with a finite spin projection. Such pairs are turned into low energy excitations by the TI's Rashba spin-orbit coupling when their spin has a proper orientation (helicity). They can lower their energy in proportion to their momentum, and emerge as coherent bosonic excitations of spin-currents at energies inside the fermionic bandgap. A triplet condensation then becomes possible by applying a gate voltage, and it occurs at finite momenta. A ``helical'' triplet condensate is expected to host a vortex lattice of spin-currents shaped by the strong SU(2) ``magnetic'' flux of the spin-orbit coupling\cite{Nikolic2011a, Nikolic2012a} (strictly speaking, only the translation symmetry is spontaneously broken in this state, and only at zero temperature\cite{Moore1989, Tesanovic1994, Sinova2002}). The phase transition to a non-superconducting state is first-order at sufficiently low temperatures and originates in the zero-temperature vortex lattice melting by quantum fluctuations. The ensuing insulating state is strongly correlated and a candidate for novel fractional incompressible liquids with non-Abelian quasiparticles\cite{Nikolic2011, Nikolic2012}. All these correlated states of triplet Cooper pairs can coexist in the TI with a singlet condensate.

When a gate voltage $V_{\textrm{g}}$ is applied across the device to attract electrons from the SC to the TI, the band-structure of the bulk spectrum assumes the qualitative shape shown in the Fig.\ref{SCTI}. If the TI material were thick, its surfaces would host a topologically protected band of massless Dirac quasiparticles with a helical locking between their spin and momentum. However, the quantum well is thin enough to hybridize these surface states on its opposite sides, so a two-dimensional bandgap $2\Delta$ opens and the surface Dirac quasiparticles become massive. The hybridized surface states still live at energies well below the TI's bulk bands, and their bandgap can be of the order of $\Delta \sim 10 - 100 \textrm{ meV}$. The SC's spectrum consists of Bogoliubov quasiparticles whose (pairing) gap $\Delta_0 \sim 1 \textrm{ meV}$ is set by the critical temperature $T_c \sim 10 \textrm{ K}$. Charge is very mobile inside the SC, so this layer (the material C) provides a particle reservoir for the TI quantum well (the material UC) and fixes the chemical potential $\mu$ shown by the dashed line in the Fig.\ref{SCTI}. The same chemical potential must be established in the TI because the interface and the tunneling barrier do not prevent the transfer of particles. The gate voltage then controls the relative energy between the TI's and SC's electrons, or equivalently the position of the TI's chemical potential relative to its bands. Since only a fraction $\Delta V \sim V_{\textrm{g}} (l_\textrm{B} + d)/(l_\textrm{B} + d + l_\textrm{I})$ of the applied voltage affects the quantum well, it is desirable that the insulator thickness $l_\textrm{I}$ be as small as possible without jeopardizing the insulating behavior of I. Note that the two-dimensional character of the TIs bands is protected in the range of energies $\mu \pm \Delta_0$ where there are no bulk SC states that could resonantly couple to the TI across the interface.

This device has an interesting application described earlier only if the chemical potential can be effectively moved in the vicinity of a surface band edge in the TI. A generic solution would be to engineer a sufficiently thick tunneling barrier that can support voltage drops of the order of the bulk bandgaps ($\sim 0.3 - 0.5 \textrm{ eV}$) without suffering an electrical breakdown. This would, unfortunately, ruin the proximity effect. Instead, a barrier $l_{\textrm{B}} \sim 10 \textrm{ \AA}$ thin could be theoretically used to tune the chemical potential in the TI within a range of a few or at most ten $\textrm{meV}$ (using AlGaAs as an example). This would be quite enough to drive the subtle quantum phase transitions of triplet Cooper pairs, but still requires that the materials' interfaces in the heterostructure naturally put the TI's chemical potential within about $10 \textrm{ meV}$ from one of its hybridized band edges. Therefore, a careful choice of materials, doping, etc. may be needed to obtain the adequate work functions and other interface properties. The interaction proximity effect might still be felt a short distance $l_{\textrm{B}}$ away, helped by the fact that an insulating barrier cannot screen the charge of the SC's electrons and atoms. A tunneling barrier could be even unnecessary, and removing it would be the best for proximity effect. Namely, the \emph{insulating} TI can itself play the role of a tunneling barrier for the purposes of creating inter-orbital spin triplet Cooper pairs. Its desired thickness of the order of $d \sim 10 \textrm{ \AA}$ is likely sufficient to tune the chemical potential across the low-energy landscape of possible correlated phases. This characteristic range of energies for correlated states is bounded by the SC's critical temperature, which is at best of the order of $1 \textrm{ meV}$.

Here, however, our goal is not to investigate the microscopic device design, but rather the device's fundamental physical feasibility and principles of its operation. We will particularly show that a sufficiently strong interaction proximity effect can create a triplet bound-state Cooper pair in the TI, whose band lives below the fermionic quasiparticle continuum. The bound-state is made possible by the two-dimensional geometry of the quantum well\cite{L1977} and its insulating ground state. Furthermore, the triplet bound states with large net momentum can gain energy from their Rashba spin-orbit coupling and become even more advantageous than singlets. Putting the chemical potential in their low-energy band creates a triplet ``condensate'' (in the mean-field approximation), which in this case hosts the TR-invariant vortex lattice mentioned earlier.

The physics portrayed here should be contrasted with other examples of proximity effects and two-dimensional superconductivity. First of all, the proximity effect applied to a 2DES with no orbital degree of freedom, such as the surface of a bulk TI, can hardly produce a triplet Cooper pair of sufficiently small size whose spin-orbit coupling could turn it into a competitor to singlets for condensation. Furthermore, no Cooper pair could exist as a bound state when the fermion quasiparticle spectrum is gapless. In contrast, the system in the Fig.1 provides an orbital degree of freedom (surface index) to electrons, confines electrons in the $z$ direction to allow small inter-orbital triplet pairs, and maintains an energy gap in the quasiparticle spectrum which can be partially filled by bosonic excitations. This is the basis for the existance of novel correlated states in this system. Should a non-Abelian fractional TI be realized in this system, it might provide a more robust platform for topological quantum computation than the Majorana quasiparticle route. Namely, Majorana quasiparticles remain topologically protected only while being kept a large distance apart, while the non-Abelian quasiparticles of a fractional TI are naturally gapped and keep their mutual fractional statistics at finite distances.

A Cooper pair bound state is not normally encountered in condensed matter systems. Its existence here is tied to the ``band-insulating'' state of the \emph{two-dimensional} TI, where any amount of (induced) attractive potential between particles creates a bound-state\cite{L1977, R1989, SR1989, Nussinov05}. This pre-formed Cooper pair still has quite a large size, so it quickly loses its identity in the correlated finite-density ground states. The main consequence of the pair binding energy is that strongly correlated quantum states at low-densities are possible, which is particularly important for the goal of obtaining fractional incompressible liquids (where the density of particles should be of the order or smaller than the density of flux quanta). In other words, the insulator-superconductor phase transition (of spinful triplets) is not merely a pairing transition, but a less conventional one: first-order, or belonging to the bosonic mean-field universality class if second-order\cite{nikolic:144507, Nikolic2010, Nikolic2010b, Nikolic2011a}. One of our main results is that the Cooper pair binding energy, which sets the scale for the correlated phases of interest, need not be much smaller than the SC's critical temperature scale $T_c \sim 10 \textrm{ K}$. Thus, the novel interesting correlated states should be experimentally accessible in the system of this kind.

The interaction proximity effect may produce unconventional correlated states of electrons in a variety of systems, including non-topological materials. The nature of any such correlated states depends on the intrinsic materials' dynamics, and their classification goes beyond the scope of this paper. Our discussion repeatedly refers to only one concrete system, the TI quantum well, because its intrinsic dynamics may support topological triplet pairing with different correlations than those found in the proximate conventional superconductor. The essential features of electron dynamics that stimulate triplet pairing are internal ``orbital'' degrees of freedom with (nearly) degenerate energies near the interface, and a strong spin-orbit coupling. Both are naturally present in the TI quantum wells modeled by (\ref{Bernevig}), but could be featured in some other materials or engineered interfaces as well. Our subsequent analysis of the interaction proximity effect will not make any concrete assumptions about the degrees of freedom available in the affected 2DES, or their dynamics. We will only determine the ways in which the SC's degrees of freedom (lattice phonon modes and BCS-paired electrons) renormalize or induce the interactions among the 2DES electrons.

\subsection{The superconducting proximity effect}\label{secProxi}

Our discussion will focus on the interfaces between \emph{conventional} superconductors and TIs, following the current experimental efforts \cite{Kasumov1996, Koren2011, Qu2011, Sacepe2011, Yang2011, Zhang2011, Veldhorst2012, Wang2012, Williams2012}. There are two sources of effective attractive interactions among electrons in the TI near the interface: the phonon mechanism and virtual Cooper pair tunneling. The former is the same microscopic BCS mechanism that is responsible for pairing in the SC material. Any charge-carrying excitation of the TI's surface electrons can displace the atoms of the SC crystal near the interface, and thus attractively interact with another TI's excitation via emission and absorption of the SC's phonons. The Coulomb origin of this interaction makes it spin-independent and capable of creating triplet Cooper pairs in principle. Its range is either short due to screening, or blunted by geometry (we describe this in detail in the section \ref{secCoulomb}). This attractive interaction is further aided by the phonons of the remaining TI's environment.

The second source of effective interactions is the dynamics of the SC's Cooper pairs. Electrons from the SC can gain kinetic energy by tunneling into the TI and back, provided that they can maintain their pairing correlation while being inside the TI. Therefore, tunneling can dynamically generate pairing forces in the TI, which are short-ranged due to the SC's Meissner effect. These dynamic forces reflect the SC's pairing and thus are spin-dependent, but we will show that they nevertheless also contribute to the possibility of triplet pairing. Cooper pair tunneling also produces the conventional proximity effect, a direct imprint of the SC's $s$-wave order parameter across the interface. This is important in a metallic surface 2DES, but becomes only a small perturbative effect in the TI \emph{quantum wells} of our interest (the explicitly induced pairing gap is $\sim \Delta_0^2 / \Delta \ll \Delta_0$).

The total generated attractive interaction in the TI has to be sufficiently strong to overcome the Coulomb repulsion and form Cooper pairs. Fortunately, it seems that the needed conditions for pairing and superconductivity can be achieved in the gated device from the Fig.\ref{SCTI}. Pairing in the inter-orbital spin-triplet channels, between two electrons of equal spin on the opposite quantum well surfaces, is also feasible. Namely, the Rashba spin-orbit coupling lowers the energy of triplet pairs, and the Coulomb forces that the phonon mechanism is based on are unscreened or poorly screened across the quantum well. Some in-plane screening inside the TI is always provided by the SC's electrons via the image-charge effect, but screening across a thin quantum well cannot be effective even in the proximity-induced superconducting state because the TI's Cooper pair condensate is fundamentally two-dimensional so its wavefunction is rigidly determined in the $z$-direction by the quantum well confinement. The dynamics of some TI's electron's remains two-dimensional so they can form bound-state Cooper pairs.

\subsection{The contents and main results}

The first part of the paper, section \ref{secPhonons}, analyzes the phonon-mediated interactions. We will first review in the section \ref{sec2DBCS} the BCS theory prediction that the pairing gap is not a function of the underlying Fermi surface size in two-dimensions. This is in stark contrast to three-dimensional superconductors. We will show in the section \ref{secBound} that this result naively extrapolates all the way to the zero density, where the Fermi surface is shrunk and disposed of in favor of a band-insulating ground state. We will recall another well-known result that a bound-state always exists in two-dimensional attractive potentials, and show that the binding energy has the same dependence on the electron-phonon coupling as the 2D superconductor's pairing gap. This calculation is simple, but ignores the phonon dynamics. We will redo the calculation in the section \ref{secBetheSalpeter} using the Bethe-Salpeter equation to fully take into account the phonon dynamics, and argue that the Cooper pair bound-state remains stable despite retardation effects. Finally, we will analyze the fate of the bound-state against the Coulomb repulsion in the section \ref{secCoulomb} and show that it survives when the electron-phonon coupling is large enough. The bound-state criterion is a slight geometric modification of the analogous criterion for the existance of superconductivity in the 3D bulk SC material. Based on this we argue that a sufficiently thin quantum well (and tunneling barrier) should allow the phonon-mediated pairing in realistic SC-2DES interfaces. The binding energy estimates in this paper do not include any effects of the spin-orbit couplings. Such effects have been taken into account elsewhere\cite{Nikolic2012a}, and only help the triplet Cooper pairs to form a two-dimensional band within the fermionic bandgap.

In the second part of the paper, section \ref{secTunneling}, we consider the tunneling effect of Cooper pairs on the 2DES. We model the interface simply as a tunneling barrier and derive the effective action of the 2DES electrons by integrating out the SC's electron degrees of freedom using a path-integral. The section \ref{secSinglet} explores the simplest consequences of tunneling where only the shortest-range interactions are induced. The effective action contains a direct symmetry-breaking coupling between the 2DES electrons and the SC's order parameter (the conventional proximity effect), and also an effective interaction term. The latter contributes to the same type of correlation as the conventional proximity effect. In the following section \ref{secTriplet} we analyze the tunneling-generated interactions in any available inter-orbital channels, which also includes spin-triplet channels. We find that such attractive interactions are generally induced by quantum fluctuations, but probably too weak to bring about any new physics on their own. They may, however, help the phonon mechanism to generate triplet pairing. Aspects of Coulomb interactions between the 2DES and SC electrons are covered in the section \ref{secCoulomb}.

\section{Phonon mechanism}\label{secPhonons}

Here we focus on the phonon-mediated proximity effect in a band-insulator quantum well that touches a conventional superconductor. We will argue that the phonon mechanism can give rise to bound-state Cooper pairs in the two-dimensional band-insulator, whose binding energy scale may be large enough to observe. This scale determines the critical chemical potential (away from the electron's conduction or valence band) for the zero-temperature superconductor-insulator transition, as well as the typical critical temperature deeper in the superconducting phase. The analysis has four parts that gradually build a full picture of the proximity phonon mechanism and eventually take the Coulomb repulsion into account.

The coupling between electrons and phonons was studied in great detail for a long time \cite{Mahan2000}. The minimal model of electron and phonon dynamics, which contains all important ingredients for conventional superconductivity, is given by the Hamiltonian
\begin{eqnarray}
H \!\! &=& \!\! \int \dd^d r \biggl\lbrack
  \frac{1}{2m} (\boldsymbol\nabla \psi_\sigma^\dagger) (\boldsymbol\nabla \psi_\sigma^{\phantom{\dagger}})
  -\mu \psi_\sigma^\dagger \psi_\sigma^{\phantom{\dagger}} \nonumber \\
&& + ug \, \psi_\sigma^\dagger \psi_\sigma^{\phantom{\dagger}} \boldsymbol{\nabla\phi}
   + \frac{1}{2} \biggl( {\dot{\boldsymbol{\phi}\phantom{i}}\!\!}^2 + u^2 (\boldsymbol{\nabla\phi})^2 \biggr) \biggr\rbrack
\end{eqnarray}
in the continuum limit. The real vector field $\boldsymbol{\phi}$ describes the ``displacement density'' of atoms in the crystalline lattice, and $\dot{\boldsymbol{\phi}\phantom{i}}\!\!$ is its canonical conjugate. $\psi_\sigma$ is the electron field (annihilation) operator. This Hamiltonian is valid at energy and momentum scales where the dispersion of sound is approximately determined by a single parameter, the sound velocity $u\sim 10^3\textrm{ m/s}$. The point-group symmetries forbid the minimal linear electron density coupling to transverse phonons. Any transverse coupling must involve either scalar combinations of $\boldsymbol{\phi}$ or electron currents, so it can be neglected perturbatively due to the smallness of its coupling constant or electromagnetic relativistic effects. The electron $G$ and longitudinal phonon $D$ propagators are given by the time-ordered ground-state expectation values:
\begin{eqnarray}\label{Prop}
G_{\sigma\sigma'}(x_\mu^{\phantom{k}}, x_\mu') \!\! &=& \!\! -i \langle 0|
  T \psi_\sigma^{\phantom{k}}(x_\mu^{\phantom{k}}) \psi_{\sigma'}^\dagger(x_\mu') |0 \rangle \\
D(x_\mu^{\phantom{k}}, x_\mu') \!\! &=& \!\! -i u^2 \langle 0|
  T \bigl\lbrack\boldsymbol{\nabla\phi}(x_\mu^{\phantom{k}})\bigr\rbrack
    \bigl\lbrack\boldsymbol{\nabla\phi}^\dagger(x_\mu')\bigr\rbrack |0 \rangle \nonumber \ ,
\end{eqnarray}
and their Fourier transforms are:
\begin{eqnarray}\label{EPprop}
G_{\sigma\sigma'}(p_\mu) \!\! &=& \!\! \frac{\delta_{\sigma\sigma'}}{\omega-E_{\bf p}+i0^+\textrm{sign}(E_{\bf p})} \\
D(p_\mu) \!\! &=& \!\! \frac{u^2 |{\bf p}|^2}{\omega^2 - u^2 |{\bf p}|^2 + i0^+} \ . \nonumber
\end{eqnarray}
The electron-phonon coupling $g$ has been associated with the vertex. We denote by $x_\mu = ({\bf r},t)$ and $p_\mu = ({\bf p}, \omega)$ the space-time positions and momenta respectively (using the units $\hbar=1$), and $E_{\bf p} = |{\bf p}|^2/2m-\mu$ is the electron energy relative to the chemical potential $\mu$.

\subsection{Two-dimensional BCS superconductors}\label{sec2DBCS}

\begin{figure}
\includegraphics[height=0.75in]{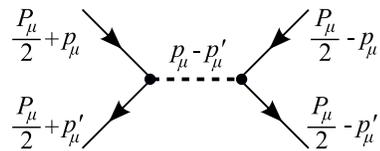}
\caption{\label{ees}Phonon-mediated scattering between two electrons.}
\end{figure}

The BCS theory of superconductivity in metals relies on several approximations to predict the properties of the superconducting state. First, the electron-phonon coupling $g$ is assumed to be small, so vertex corrections can be neglected. Second, phonon-mediated energy transfers between interacting electrons can be neglected on average. This is justified in metals with sufficiently large electron density: typical momentum transfers among the dynamically active electrons near the Fermi surface are of the order of Fermi momentum $p_{\textrm{f}} = \sqrt{2m E_{\textrm{f}}} = m v_{\textrm{f}}$, where $v_{\textrm{f}}\sim 10^6\textrm{ m/s} \gg u$, while the maximum energy transfer by phonons is of the order of the Debye frequency $\omega_{\textrm{D}} \sim u p_{\textrm{f}} \ll E_{\textrm{f}}$. The electron-electron scattering process represented by the Feynman diagram in Fig.\ref{ees} can then be approximated by an attractive interaction via a static instantaneous contact potential of strength $\lambda_0 = g^2$:
\begin{equation}\label{Ueff}
U({\bf p},\omega) = g^2 D({\bf p},\omega) = g^2 \frac{u^2|{\bf p}|^2}{\omega^2-u^2|{\bf p}|^2} \approx -\lambda_0 \ ,
\end{equation}
because $|\omega| \ll u|{\bf p}| \sim \omega_D$ for typical phonons. In the last mean-field-like approximation, the pairing gap is determined by assuming that all electrons near the Fermi surface contribute equally to superconductivity as long as they are capable of exchanging energy up to $\omega_{\textrm{D}}$ with other electrons. The pairing gap $\Delta_0$ at zero temperature is accordingly calculated from the self-consistent condition\cite{Abrikosov1975}:
\begin{equation}\label{SSgap}
\frac{\lambda_0}{2}\int\frac{\dd^d p}{(2\pi)^d}\,\frac{\theta(\omega_{\textrm{D}}-|E_{\bf p}|)}{\sqrt{E_{\bf p}^2+\Delta_0^2}}=1 \ ,
\end{equation}
and leads to the well-known result in three-dimensional ($d=3$) metals:
\begin{equation}\label{Gap3D}
\Delta_0 \approx 2\omega_{\textrm{D}} \exp\left(-\frac{2\pi^2}{\lambda_0 m \, p_{\textrm{f}}}\right) \ll \omega_{\textrm{D}} \ .
\end{equation}
The same energy scale also determines the critical temperature $T_{\textrm{c}} \sim \Delta_0$.

Consider now a hypothetical two-dimensional weak-coupling superconductor. The pairing gap is now determined from (\ref{SSgap}) with a two-dimensional electron-phonon coupling $\lambda$:
\begin{eqnarray}
\frac{\lambda}{2}\!\int\!\frac{\dd^2 p}{(2\pi)^2}\,
  \frac{\theta(\omega_{\textrm{D}}-|E_{\bf p}|)}{\sqrt{E_{\bf p}^2+\Delta_{2\textrm{D}}^2}}
\!\! &\approx& \!\! \frac{\lambda}{2}\frac{2\pi p_{\textrm{f}}}{(2\pi)^2 v_{\textrm{f}}} \int\limits_{-\omega_{\textrm{D}}}^{\omega_{\textrm{D}}}
  \!\!\frac{\dd\epsilon}{\sqrt{\epsilon^2+\Delta_{2\textrm{D}}^2}} \nonumber \\
\!\! &\approx& \!\! \frac{\lambda m}{2\pi} \log\left(\frac{2\omega_{\textrm{D}}}{\Delta_{2\textrm{D}}}\right) = 1 \ .
\end{eqnarray}
Therefore,
\begin{equation}\label{Gap2D}
\Delta_{2\textrm{D}} \approx 2\omega_{\textrm{D}}\exp\left(-\frac{2\pi}{\lambda m}\right) \ll \omega_{\textrm{D}}
\end{equation}
in two dimensions. The striking difference from (\ref{Gap3D}) is that there is no dependence on the Fermi momentum $p_{\textrm{f}}$. In three dimensions it is important to have a large Fermi surface with a large $p_{\textrm{f}}$, otherwise $\Delta_0$ and the corresponding critical temperature can be exponentially suppressed to an unobservable small value. In contrast, the pairing gap and critical temperature in two dimensions do not depend on the electron density as long as the approximations behind this simple model are valid. Instead of extrapolating this result down to zero density, we will consider more appropriate models in the following sections.

The main system of our interest is a quantum well interfaced with a conventional superconductor. The well could resemble a BCS-paired two-dimensional metal if a fairly large gate voltage were applied to create a dense two-dimensional electron gas in it. However, we will shortly turn our attention to insulating wells. A very thin quantum well is not likely to contribute significant longitudinal phonons of its own, and we will not consider its transverse phonons either. The latter could be minimally but weakly coupled because the inversion symmetry is absent in the $z$-direction. The superconductor's phonons may anyway be the most likely ones to have a significant coupling to the electrons in the quantum well. This coupling $\lambda$ is surely smaller, but need not be much smaller than the superconductor's intrinsic coupling adjusted for the different dimensionality of the quantum well. We will therefore loosely estimate $\lambda \sim \lambda_0 d^{-1}$, where $d$ is the quantum well thickness, or $\lambda \sim \lambda_0 (d+l_{\textrm{B}})^{-1}$ if a tunneling barrier is present as in the Fig.\ref{SCTI}. This is expected to be valid when the quantum well and tunneling barrier are very thin.

The estimate $\lambda \sim \lambda_0 d^{-1}$ is fully justified merely by dimensional analysis if no length-scales other than $d$ characterize the static electrons in the quantum well. Specifically, it holds when the quantum well is insulating, so the density of its screening-capable electrons is zero. It is also hard for the superconductor's electrons to tunnel into an insulating quantum well. Hence, they cannot screen the Coulomb interaction between the surface ions and any excited quantum well electrons, but only reduce $\lambda$ by a factor of order one via the image-charge effect. Further reduction of $\lambda$ by a factor of order one can come from the modified elastic properties of the crystals near the interface. However, this effect is expected to be small. The superconductor's elastic crystal may be interfaced with an effectively rigid medium in the worst-case scenario, which then provides a fixed boundary condition for the displacements of the superconductor's surface atoms. Only the shortest-wavelength phonons are affected by this, which can be approximately modeled by a suppressed phonon density of states above some cut-off momentum (in the $z$-direction). Our calculations will eventually include such a cutoff, but the simple truth is that the typical phonon-mediated momentum transfers between \emph{insulating} quantum well's electrons are small.

Therefore, the critical temperature $T_{\textrm{c}}^{\textrm{2D}}$ of the ensuing two-dimensional superconductivity could be comparable to that of the three-dimensional superconductor $T_{\textrm{c}} \sim 10 \textrm{ K}$ at least if the tunneling barrier is removed. Even a more realistic reduction of $T_{\textrm{c}}^{\textrm{2D}}$ by an order of magnitude or two (due to its exponential dependence on $\lambda$) can still be experimentally observable.

\subsection{A simple model of Cooper pair bound states}\label{secBound}

The previous section has suggested that a superconducting state can survive in two-dimensional metals at any density of electrons. What if the density is lowered to zero? Is it possible to have a phase transition between a band-insulator and a paired superconductor across which the fermion excitation gap remains finite?

The renormalization group argument yields a positive answer to this question in two dimensions, for any strength of attractive interactions\cite{Nikolic2010}. This is a consequence of the fact that any short-range attractive potential always produces a two-body bound state in two dimensions\cite{L1977}. A two-dimensional Cooper pair ``molecule'' may be very large and fragile, but it always has a finite binding energy. If their Bose condensation is arranged from the band-insulating state, the second order phase transition belongs to the mean-field bosonic universality class and hence differs from the standard pairing BCS transition. The normal state adjacent to this transition is a ``pseudogap'' bosonic Mott insulator, since its lowest energy excitations are gapped Cooper pairs lying inside the fermion bandgap. In the most mundane circumstances, the Mott insulator is separated from the band-insulator in the phase diagram only by a crossover. However, if the transition involves triplet Cooper pairs, or inter-valley pairing, interesting correlated Cooper pair insulators that break some symmetry become a possibility in two dimensions. In the case of a vortex lattice state predicted to result from triplet superconductivity in topological insulator quantum wells, the above second order transition is preempted by the first order vortex lattice melting that potentially stabilizes a fractional topological insulator\cite{Nikolic2011, Nikolic2012}.

It is this range of interesting possibilities that motivates us to explore the bound-state Cooper pairs. The binding energy sets the temperature scale below which the correlated insulators and low-density superconductors can be observed in the quantum well. Our main goal is to estimate this energy scale and determine if it can be reasonably large. For simplicity, we will consider only the pairing of standard non-relativistic electrons. More interesting pairing in the presence of a strong Rashba spin-orbit coupling also reveals a finite binding energy\cite{Nikolic2012a}.

We will here without any justification neglect energy transfers by phonons in two-dimensional band-insulators, and approximate the phonon-mediated scattering between electrons by a short-range potential with strength $\lambda$. After getting a crude estimate of the pair binding energy, we will consider the effect of energy transfers in the next section. The effective second-quantized Hamiltonian in the quantum well derived from the phonon mechanism without energy transfers is:
\begin{equation}
H = \int \dd^2 r \biggl\lbrack
  \frac{1}{2m} (\boldsymbol\nabla \psi_\sigma^\dagger) (\boldsymbol\nabla \psi_\sigma^{\phantom{\dagger}})
  +\epsilon_0 \psi_\sigma^\dagger \psi_\sigma^{\phantom{\dagger}}
  -\lambda \psi_{\sigma}^\dagger \psi_{\sigma'}^\dagger \psi_{\sigma'}^{\phantom{\dagger}} \psi_{\sigma}^{\phantom{\dagger}}
    \biggr\rbrack \ .
\end{equation}
We are suppressing any orbital indices that the electron fields may have, and allowing $\sigma = \sigma'$ if an orbital degree of freedom exists. The chemical potential $\mu$ has been replaced by an excitation gap $\epsilon_0>0$ appropriate for a band-insulator. Since there are no particle excitations in the ground-state, we can solve the two-body problem using the equivalent single-quantized Schrodinger equation:
\begin{equation}
\left\lbrack \frac{p_1^2}{2m}+\frac{p_2^2}{2m}+2\epsilon_0-\lambda\delta({\bf r}_1-{\bf r}_2)\right\rbrack \Psi({\bf r}_1,{\bf r}_2)
  = E\Psi({\bf r}_1,{\bf r}_2) \ .
\end{equation}
By separating the center-of-mass from the relative motion $\Psi({\bf r}_1,{\bf r}_2) = \psi_{\textrm{cm}}({\bf R}) e^{il\theta} \psi(r)$, where ${\bf R} = ({\bf r}_1+{\bf r}_2)/2$ and ${\bf r} = {\bf r}_1 - {\bf r}_2 = (r,\theta)$ in cylindrical coordinates, we obtain the reduced radial Schrodinger equation in the center-of-mass frame for $\psi(r)$:
\begin{equation}\label{Schr2D}
-\frac{1}{m}\frac{1}{r}\frac{\partial}{\partial r}\left(r\frac{\partial\psi}{\partial r}\right)+\frac{l^{2}}{m r^2}\psi
  -\lambda \frac{\theta(a-r)}{\pi a^2} \psi = \epsilon\psi \ .
\end{equation}
The reduced mass of the relative motion is taken into account, and $\epsilon = E-2\epsilon_0$ is the Cooper pair binding energy if negative. At this point we have introduced an ultra-violet cutoff length $a$ and regularized the Dirac delta-function in the contact potential:
\begin{equation}
\delta({\bf r}) = \lim_{a \to 0} \frac{\theta(a-r)}{\pi a^2} \ .
\end{equation}
This is necessary in order to solve the equation (\ref{Schr2D}). The scale $a$ is defined by the momentum cutoff $\Lambda = a^{-1}$ for the phonon propagator in (\ref{Prop}). The acoustic phonon-mediated interaction between electrons cannot be significant at electron separations smaller that the lattice constant, so we expect $a$ to be of the order of lattice spacing.

This is a simple and standard problem in quantum mechanics. The solution for $r>a$ is expressed in terms of the Bessel functions:
\begin{equation}\label{WF1}
(\forall r>a)\quad\psi(r)=C_{1}J_{l}(k'r)+C_{2}Y_{l}(k'r)\quad,\quad k'=\sqrt{M\epsilon} \ .
\end{equation}
The energy $\epsilon$ of a bound state is negative, so that the parameter $k' = i\kappa$ is imaginary and the Bessel functions exhibit an exponential dependence on $r$. Their asymptotic behavior in the $r\to\infty$ limit
\begin{eqnarray}
&& J_l(i\kappa r)\xrightarrow{r\to\infty}\sqrt{\frac{1}{2\pi\kappa r}}e^{\frac{i\pi}{2}l}e^{\kappa r} \nonumber \\
&& Y_l(i\kappa r)\xrightarrow{r\to\infty}i\sqrt{\frac{1}{2\pi\kappa r}}e^{\frac{i\pi}{2}l}e^{\kappa r} \nonumber
\end{eqnarray}
implies that we must choose $C_2 = iC_1$ in order to have a normalizable wavefunction. Inside the area affected by attraction the wavefunction is oscillatory:
\begin{equation}
(\forall r<a)\quad\psi(r)=A J_l(kr)\quad,\quad k=\sqrt{m\left(\frac{\lambda}{\pi a^2}-|\epsilon|\right)} \ .
\end{equation}
We must match the wavefunction values and derivatives on the left $r=a-0^+$ and right $r=a+0^+$ side of the potential step. We will do that only for $l=0$ because the existance of bound states is not guarantied when the angular momentum $l$ is finite and $a$ is small enough (definitely the case if $m \lambda<\pi$). We shall further anticipate that the binding energy $|\epsilon|$ satisfies
\begin{equation}\label{Kappa}
\kappa a \ll 1 \quad,\quad \kappa=\sqrt{m|\epsilon|}
\end{equation}
because $a$ is small and the attractive potential $\lambda$ is weak. The approximate form of the wavefunction (\ref{WF1}) in this limit is:
\begin{equation}
\psi(r)
  \xrightarrow{r\to a+0^+}\frac{iC}{\pi}\log\left(\kappa r\right) \ ,
\end{equation}
so that the boundary conditions at $r=a$ for $l=0$ are:
\begin{eqnarray}
&& AJ_0(ka) = \frac{iC}{\pi}\log(\kappa a) \\
&& \frac{Ak}{2}\Bigl(J_{-1}(ka)-J_{1}(ka)\Bigr) = \frac{iC}{\pi a} \ . \nonumber
\end{eqnarray}
Dividing these two equations yields an energy quantization condition:
\begin{equation}
-\frac{J_0(ka)}{J_1(ka)}=ka\,\log(\kappa a) \ .
\end{equation}
By the approximation (\ref{Kappa}),
\begin{equation}
ka=\sqrt{\frac{m\lambda}{\pi}-(\kappa a)^2}\approx\sqrt{\frac{m\lambda}{\pi}}
\end{equation}
so that:
\begin{eqnarray}\label{BindingEnergy}
&& \epsilon = -\frac{\kappa^2}{m} \approx
  -\frac{1}{ma^2}\exp\left(-2\sqrt{\frac{\pi}{m\lambda}}\,\frac{J_0(\sqrt{m\lambda/\pi})}{J_1(\sqrt{m\lambda/\pi})}\right)
\nonumber \\ && ~~~~~ \xrightarrow{m\lambda\ll 1} -\frac{1}{ma^2}\exp\left(-\frac{4\pi}{m\lambda}\right) \ .
\end{eqnarray}
The final result applies in the limit $\kappa a \ll m\lambda \ll 1$ and should be compared with the pairing gap (\ref{Gap2D}) of the two-dimensional BCS superconductor. The dependence on the electron-phonon coupling $\lambda$ is essentially the same, which justifies the earlier naive assumption that (\ref{Gap2D}) can be extrapolated to very low densities. The difference between the exponents of (\ref{Gap2D}) and (\ref{BindingEnergy}) can be attributed to the effective mass reduction in relative motion, which is not taken into account in (\ref{Gap2D}). The Cooper pair size $\xi\sim\kappa^{-1}\gg a$ is relatively large.

The energy scale $1/(ma^2)$ obtained here can be much larger than $\omega_{\textrm{D}}$, but depends on the microscopic details at the cutoff scale which we cannot model accurately. This result has only a qualitative significance. It reflects the fact that pairing is much more efficient in two dimensions than in three dimensions. The strict weak-coupling limit of pairing does not even exist in two-dimensional systems at very low densities \cite{Nikolic2010, Nikolic2010b}, and superconductors with (quasi) two-dimensional dynamics tend to have higher critical temperatures (MgB$_2$ is an example).

An even more important point to emphasize is that our  two-dimensional system of interest is not isolated as the above calculation indirectly assumes, but coupled to a bulk superconductor whose spectrum has a continuum of extended states at energies more than $\Delta_0$ away from the chemical potential $\mu$. Only within the energy range $2\Delta_0$ can the dynamics in the quantum well be regarded as two-dimensional. We will see in the following section that this cuts off the Cooper pair binding energy down to no more than the bulk superconductor's pairing gap scale $\Delta_0$:
\begin{eqnarray}\label{BindingEnergy2}
&& |\epsilon| \sim E_0 \exp\left(-\frac{4\pi}{m\lambda}\right) \quad,\quad E_0 \sim \Delta_0 \ .
\end{eqnarray}
Nevertheless, we will argue in the section \ref{secCoulomb} that this more realistic pair binding is experimentally observable in some cases.

\subsection{Cooper pair bound states from the Bethe-Salpeter equation}\label{secBetheSalpeter}

Neglecting the phonon dynamics in electron scattering is actually not justified in band-insulators or low-density metals. A typical momentum transfer can be arbitrarily small in the absence of a Fermi surface, and thus smaller than the energy transfer from one electron to another in a scattering event. Given that the speed of sound is small on the scale of typical electron velocities, retardation effects caused by the exchange of phonons could be quite large. We can take them into account only by the proper field-theoretical treatment of scattering. Therefore, we will solve here the Bethe-Salpeter equation for the 2DES to find the Cooper pair bound state. This is a relatively complex task, so we will make certain approximations to reach the important qualitative conclusions quickly. We will find that the retardation and quantum effects reduce but not eliminate the short-range attractive potential responsible for the bound-state. By considering the worst case scenarios, we will argue that the effective short-range attraction is weakened from the one used in the previous section by a factor of two or so.

\begin{figure}
\includegraphics[height=0.9in]{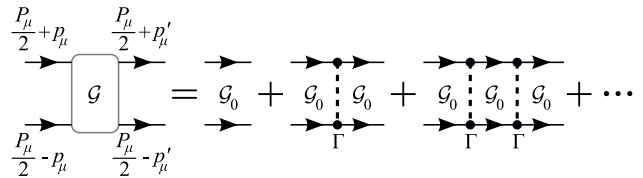}
\caption{\label{BSfig}The diagrammatic Dyson equation for the two-particle Green's function $\mathcal{G}$.}
\end{figure}

The Bethe-Salpeter equation is obtained from the Dyson equation for the two-particle Green's function, whose diagrammatic representation is shown in the Fig.\ref{BSfig}:
\begin{equation}\label{Dyson}
\mathcal{G} = \mathcal{G}_0 + \mathcal{G}_0 \Gamma \mathcal{G} \ .
\end{equation}
Technical details behind its derivation and properties are reviewed in the appendix \ref{app}. The full two-body Green's function $\mathcal{G}$, the bare Green's function $\mathcal{G}_0$ and the interaction vertex $\Gamma$ are represented by matrices whose rows and columns are indexed by $(P_\mu,p_\mu$), where $P_\mu = ({\bf P},\Omega) = p_{1\mu}+p_{2\mu}$ is the total and $p_\mu = ({\bf p},\omega) = (p_{1\mu}-p_{2\mu})/2$ the relative momentum/energy of the two electrons in a Cooper pair. If there is a two-particle bound state with total energy $\mathcal{E}_{\bf P}$, then the Green's function has a pole as a function of $\Omega$ and can be separated into a coherent and incoherent part:
\begin{eqnarray}\label{Gcoh}
&& \mathcal{G}(P_\mu^{\phantom{k}},p_\mu^{\phantom{k}};P_\mu',p_\mu') =
  \frac{\mathcal{F}({\bf P}; p_\mu^{\phantom{k}}) \mathcal{F}^\dagger({\bf P}';p_\mu')}{\Omega-\mathcal{E}_{\bf P}+i0^+} \times \\
&& ~~~~~~~~~ \times (2\pi)^3 \delta(P_\mu^{\phantom{k}}-P_\mu')
    + \mathcal{G}_{\textrm{inc}}(P_\mu^{\phantom{k}},p_\mu^{\phantom{k}};P_\mu',p_\mu') \ . \nonumber
\end{eqnarray}
The total momentum and energy are conserved, hence the $\delta(P_\mu^{\phantom{k}}-P_\mu')$ factor. This conservation law also guaranties that the function $\mathcal{F}({\bf P}, p_\mu) = \mathcal{F}_{\textrm{cm}}({\bf P}) F(p_\mu)$ from the coherent term is a product of two functions associated with the center-of-mass ($\mathcal{F}_{\textrm{cm}}$) and relative ($F$) motion of the Cooper pair's electrons. We will isolate $F(p_\mu)$ because it is related to the Fourier transform $\psi({\bf p})$ of the stationary bound-state wavefunction $\psi({\bf r})$:
\begin{equation}\label{WF2}
\psi({\bf p}) = \int\frac{\dd\omega}{2\pi} \, F({\bf p},\omega) \ .
\end{equation}
The bare two-body Green's function is given by the product of two single-particle propagators (\ref{Prop}):
\begin{eqnarray}
&& \!\!\!\! \mathcal{G}_0(P_\mu^{\phantom{k}},p_\mu^{\phantom{k}};P_\mu',p_\mu') =
   i G\left(\frac{P_\mu}{2}+p_\mu\right) G\left(\frac{P_\mu}{2}-p_\mu\right) \times \nonumber \\
&& ~~~~~~~ \times (2\pi)^3 \delta(P_\mu^{\phantom{k}}-P_\mu') \, (2\pi)^3 \delta(p_\mu^{\phantom{k}}-p_\mu')
\end{eqnarray}
and conserves the momenta and energies of both particles. The single-particle propagators should generally contain all self-energy corrections, but in this case we will ignore them because the weakness of the electron-phonon coupling justifies working at the lowest order of perturbation theory. This also means that the vertex part $\Gamma$ need not contain more than one phonon line and two electron-phonon vertices. If the phonons were able to propagate only in the two dimensions of the quantum well, their vertex part
\begin{equation}\label{Ver2D}
\Gamma_{2\textrm{D}}(P_\mu^{\phantom{k}},p_\mu^{\phantom{k}};P_\mu',p_\mu') = g^2 D(p_\mu^{\phantom{k}} - p_\mu')
  (2\pi)^3 \delta(P_\mu^{\phantom{k}}-P_\mu')
\end{equation}
would involve only the (2+1)D momentum and energy transfer $p_\mu^{\phantom{k}} - p_\mu'$ between the two scattering electrons. However, phonons can generally propagate in all three dimensions, so their propagator $D(\delta p_\mu)$ should take a (3+1)D momentum/energy transfer. The extra momentum component $q_z$ perpendicular to the quantum well is not conserved by the coupling of phonons to the quantum well electrons. We shall regard the quantum well as being very thin and limit the momentum $q_z$ by its own cutoff $\Lambda_z \sim d^{-1}$, where $d$ is the quantum well thickness. We will absorb any geometric reduction of the electron-phonon coupling into $g_0$. The in-plane momentum components are bounded by the cutoff $\Lambda \sim a^{-1}$. Phonons with sufficiently large wavelength along the $z$-direction will couple to the electrons in the quantum well in essentially a momentum-independent manner. Therefore, we can approximate the vertex part of three-dimensional phonons as:
\begin{eqnarray}\label{Ver3D}
&& \!\!\!\!\!\! \Gamma_{3\textrm{D}}(P_\mu^{\phantom{k}},p_\mu^{\phantom{k}};P_\mu',p_\mu') = g_0^2 \int \frac{\dd q_z}{2\pi}
  D({\bf p}-{\bf p}'+\hat{\bf z}q_z, \omega-\omega') \nonumber \\
&& ~~~~~~~~~~~~~~~~~~~~~~~ \times (2\pi)^3 \delta(P_\mu^{\phantom{k}}-P_\mu') \ .
\end{eqnarray}

In order to simplify the initial discussion, we will carry out most derivations assuming that the phonons have a purely two-dimensional dispersion captured by the vertex part $\Gamma = \Gamma_{2\textrm{D}}$. It will be straight-forward to switch later to the three-dimensional phonon dispersion. The Bethe-Salpeter equation for the relative motion $F(p_\mu)$ of two electrons in a Cooper pair is obtained from (\ref{Dyson}) and (\ref{Gcoh}) by taking the limit $\Omega \to \mathcal{E}_{\bf P}$ and neglecting the terms that do not diverge on the approach to the bound-state pole:
\begin{equation}\label{BS1}
F = \mathcal{G}_0\Gamma F \quad,\quad \Omega\to \mathcal{E}_{\bf P} \ .
\end{equation}
In this equation we canceled out the common factors that depend only on ${\bf P}$ and now index all vectors and matrices by the relative (2+1)D momenta $p_\mu$. The expanded form of this equation involving two-dimensional phonons is:
\begin{eqnarray}\label{BS2}
&& \!\!\!\!\!\!\!\!\!\!\! F(p_\mu) = ig^2 \, G\left(\frac{P_\mu}{2}+p_\mu\right) G\left(\frac{P_\mu}{2}-p_{\mu}\right) \\
&& ~~~~ \times \int\frac{\dd^3 p_{\mu}'}{(2\pi)^3}\,
      D({\bf p}-{\bf p}', \omega-\omega') F(p_\mu') \ . \nonumber
\end{eqnarray}
If one could drop the frequency dependence of the phonon propagator $D$, and thus neglect energy transfers among electrons in scattering events, then the existing integral over $\omega'$ of $F({\bf p}',\omega')$ on the right-hand side would produce precisely the desired stationary wavefunction $\psi({\bf p})$ according to (\ref{WF2}). Integrating out the whole equation over $\omega$ would reduce the left-hand side to $\psi({\bf p})$ as well, so one would be left with an integral form of a stationary Schrodinger equation. This $\omega$ integral would pick a pole from one of the single-particle Green's functions on the right-hand-side, and allow a straight-forward conversion of the above integral equation to the standard differential Schrodinger equation. The outcome of this procedure would be precisely the equation we solved in the previous section. However, we will have to adapt this procedure to the case of non-negligible energy transfers, where strictly speaking the ordinary Schrodinger equation is not an accurate description of dynamics. The Schrodinger equation can describe only the stationary energy levels, but not the phonon-mediated transitions between them which are captured by the Bethe-Salpeter equation. The only bound state of a Cooper pair that remains stable and stationary despite energy transfers is the ground state. Therefore, we may sensibly reduce the Bethe-Salpeter equation to an effective Schrodinger equation only for the ground state. This is sufficient for the purpose of estimating the binding energy. The reduction will incorporate all corrections to the ground state due to the zero-point quantum fluctuations of phonons (we will assume that there is no external source of phonons, and work at zero temperature).

We are interested in the Cooper pair ground state at rest. Therefore, let us substitute the propagators (\ref{EPprop}) and ${\bf P} = 0$ in (\ref{BS2}):
\begin{eqnarray}\label{BS3}
F(p_{\mu}) \!\!&=&\!\! \frac{i g^2}
  {\left\lbrack\frac{\epsilon}{2}-\frac{|{\bf p}|^2}{2m}+\omega+i0^+\right\rbrack \!
   \left\lbrack\frac{\epsilon}{2}-\frac{|{\bf p}|^2}{2m}-\omega+i0^+\right\rbrack} ~~~~ \\
&& \!\!\!\! \times \int\frac{\dd^3 p_{\mu}'}{(2\pi)^3}\,\frac{u^{2}|{\bf p}-{\bf p}'|^2}{(\omega-\omega')^2
  -u^2|{\bf p}-{\bf p}'|^2+i0^+} F(p_{\mu}') \ . \nonumber
\end{eqnarray}
The electron chemical potential $\mu$ from $E_{\bf k}$ in (\ref{EPprop}) is negative in the band-insulating state, meaning that it takes a finite energy $|\mu|$ to create a quasiparticle excitation. We have introduced here the binding (potential) energy $\epsilon = \mathcal{E}_{\bf P} - |{\bf P}|^2/(4m) - 2|\mu| < 0$ of a Cooper pair with total energy $\Omega = \mathcal{E}_{\bf P}$.

The structure of (\ref{BS3}) suggests that we should seek a solution of the following form:
\begin{equation}\label{BSsol}
F({\bf p},\omega) = \frac{i\left(\epsilon-\frac{|{\bf p}|^2}{m}\right) \Phi({\bf p},\omega)}
  {\left\lbrack\frac{\epsilon}{2}-\frac{|{\bf p}|^2}{2m}+\omega+i0^+\right\rbrack \!
   \left\lbrack\frac{\epsilon}{2}-\frac{|{\bf p}|^2}{2m}-\omega+i0^+\right\rbrack} \ ,
\end{equation}
where $\Phi({\bf p},\omega)$ is a smooth function of $\omega$ without singularities in the upper complex half-plane. The numerator is chosen to cancel out the residue of the frequency integral in (\ref{WF2}) and establish a simple relationship to the wavefunction:
\begin{equation}
\psi({\bf p}) = \Phi\left({\bf p},\frac{\epsilon}{2}-\frac{|{\bf p}|^2}{2m}\right) \ .
\end{equation}
We will thus set $\omega = \epsilon/2-|{\bf p}|^2/2m$ and substitute the above form of $F({\bf p},\omega)$ in (\ref{BS3}):
\begin{widetext}
\begin{eqnarray}\label{BS4}
&& \!\!\!\!\!\!\! \left(\epsilon-\frac{|{\bf p}|^2}{m}\right)\psi({\bf p}) =
  g^2 \int\frac{\dd^3 p_\mu'}{(2\pi)^3}\,
  \frac{u^2|{\bf p}-{\bf p}'|^2}{(\omega-\omega')^2-u^2|{\bf p}-{\bf p}'|^2+i0^+} \,
    \frac{i\left(\epsilon-\frac{|{\bf p}'|^2}{m}\right)\,\Phi({\bf p}',\omega')}
         {\left\lbrack\frac{\epsilon}{2}-\frac{|{\bf p}'|^2}{2m}+\omega'+i0^+\right\rbrack
          \left\lbrack\frac{\epsilon}{2}-\frac{|{\bf p}'|^2}{2m}-\omega'+i0^+\right\rbrack}
    \Biggr\vert_{\omega = \frac{\epsilon}{2}-\frac{|{\bf p}|^2}{2m}} \nonumber \\[0.1in]
&& = g^2 \int\frac{\dd^2 p'}{(2\pi)^2}\,
  \frac{u^2|{\bf p}-{\bf p}'|^2}{\left(-\frac{|{\bf p}|^2}{2m}+\frac{|{\bf p}'|^2}{2m}\right)^2-u^2|{\bf p}-{\bf p}'|^2+i0^+}\,
    \psi\left({\bf p}'\right) \\
&& ~~ +\frac{g^2}{2}\int\frac{\dd^2 p'}{(2\pi)^2}\,
  \frac{u|{\bf p}-{\bf p}'|\left(\epsilon-\frac{|{\bf p}'|^2}{m}\right)}
       {\left\lbrack\epsilon-\frac{|{\bf p}'|^2}{2m}-\frac{|{\bf p}|^2}{2m}-u|{\bf p}-{\bf p}'|+i0^+\right\rbrack\left\lbrack
  -\frac{|{\bf p}'|^2}{2m}+\frac{|{\bf p}|^2}{2m}+u|{\bf p}-{\bf p}'|+i0^+\right\rbrack}\,
  \Phi\left({\bf p}',\frac{\epsilon}{2}-\frac{|{\bf p}|^2}{2m}-u|{\bf p}-{\bf p}'|\right) \nonumber
\end{eqnarray}
\end{widetext}
We have carried out the integral over $\omega'$ on the right-hand-side of (\ref{BS3}) and obtained two terms. One term is generated by the particle pole and contains the wavefunction $\psi({\bf p}')$, while the other term is generated by the phonon pole and still contains the unknown function $\Phi({\bf p}',\omega')$ at the frequency $\omega'$ that depends on the unknown energy $\epsilon$. Dealing with the latter is quite difficult, so we will introduce our first approximation. The ``phonon-pole'' integral in the last line of (\ref{BS4}) is dominated by the values of ${\bf p'}$ at which:
\begin{equation}
\frac{|{\bf p}'|^2}{2m}-\frac{|{\bf p}|^2}{2m}-u|{\bf p}-{\bf p}'| \to 0 \ .
\end{equation}
This is the only way for the integrand to diverge, given that we expect no singularities in $\Phi({\bf p}',\omega')$ and $\epsilon<0$. We will substitute the above special value of $|{\bf p}'|$ everywhere in the ``phonon-pole'' integral, except for its diverging denominator. This yields a very simplified approximate contribution of the phonon pole:
\begin{equation}
\frac{g^2}{2} \int\frac{\dd^2 p'}{(2\pi)^2}\,
  \frac{u|{\bf p}-{\bf p}'|}{-\frac{|{\bf p}'|^2}{2m}+\frac{|{\bf p}|^2}{2m}+u|{\bf p}-{\bf p}'|}\,\psi({\bf p}')
\end{equation}
whose main benefit is that the unknown $\Phi({\bf p}',\omega')$ is traded for the wavefunction $\psi({\bf p}')$ that we wish to determine. This approximation reduces the Bethe-Salpeter equation to:
\begin{widetext}
\begin{equation}\label{BS5}
\left(\epsilon-\frac{|{\bf p}|^2}{m}\right)\psi({\bf p}) \approx
  g^2 \int\frac{\dd^2 p'}{(2\pi)^2}\,\left\lbrack
    \frac{u^2|{\bf p}-{\bf p}'|^2}{\left(\frac{|{\bf p}|^2}{2m}-\frac{|{\bf p}'|^2}{2m}\right)^2-u^2|{\bf p}-{\bf p}'|^2+i0^+}
   +\frac{1}{2}\frac{u|{\bf p}-{\bf p}'|}{\left(\frac{|{\bf p}|^2}{2m}-\frac{|{\bf p}'|^2}{2m}\right)^{\phantom{2}}
     +u|{\bf p}-{\bf p}'|+i0^+}\right\rbrack \psi\left({\bf p}'\right) \ .
\end{equation}
\end{widetext}

At this point we have an integral form of a Schrodinger equation for the Cooper pair bound state, written in the momentum representation. We can easily convert it to the position representation by performing the inverse Fourier transform (note the appearance of the reduced mass $m/2$ in the kinetic energy):
\begin{equation}\label{SchrBS}
-\frac{\boldsymbol{\nabla}^2}{m}\psi({\bf r})+g^2\int \dd^2 r'\, V({\bf r},{\bf r}')\psi({\bf r}')=\epsilon\psi({\bf r}) \ .
\end{equation}
The only price to pay is that the potential $V({\bf r},{\bf r}')$ is non-local due to the retardation effects. We can view $V({\bf r},{\bf r}')$ as a non-diagonal operator in the position representation. Its components are:
\begin{equation}
V({\bf r},{\bf r}') =
  \int\frac{\dd^2 p}{(2\pi)^2}\frac{\dd^2 p'}{(2\pi)^2}\,e^{i{\bf pr}}e^{-i{\bf p}'{\bf r}'}\,\widetilde{V}({\bf p},{\bf p}') \ ,
\end{equation}
where $\widetilde{V}({\bf p},{\bf p}')$ is the expression from the square brackets of (\ref{BS5}). Upon a closer inspection of $\widetilde{V}({\bf p},{\bf p}')$ it can be easily seen that the potential energy operator $V$ is not Hermitian. The eigenvalues $\epsilon$ of the ensuing non-Hermitian Hamiltonian will generally end up being complex. This indicates that Cooper pair states have a finite lifetime of the order of $1/\textrm{Im}(\epsilon)$. Indeed, Cooper pairs can emit phonons and relax to lower energy states. Only the ground-state is stable and its eigenvalue $\epsilon$ is real. Being interested only in the ground-state, we will symmetrize the potential energy operator
\begin{equation}
V_{\textrm{s}}({\bf r},{\bf r}')=\frac{1}{2}\Bigl\lbrack V({\bf r},{\bf r}')+V^{*}({\bf r}',{\bf r})\Bigr\rbrack
\end{equation}
and hence eliminate its non-Hermitian parts. This will have no effect on the Hamiltonian projection to the ground-state, and the ground-state energy $\epsilon$. The real-space symmetrization is equivalent to:
\begin{eqnarray}\label{Vpsym}
&& \!\!\!\!\!\! \widetilde{V}_{\textrm{s}}({\bf p},{\bf p}')=\frac{1}{2}\Bigl\lbrack
   \widetilde{V}({\bf p},{\bf p}')+\widetilde{V}^{*}({\bf p}',{\bf p})\Bigr\rbrack \\
&& = \mathbb{P}\frac{u^2|\delta{\bf p}|^2}{\delta E^2-u^2|\delta{\bf p}|^2} +
     \frac{\mathbb{P}}{4}\left\lbrack
       \frac{u|\delta{\bf p}|}{\delta E+u|\delta{\bf p}|}+\frac{u|\delta{\bf p}|}{-\delta E+u|\delta{\bf p}|}
     \right\rbrack \nonumber \\
&& = \frac{\mathbb{P}}{2}\frac{u^2|\delta{\bf p}|^2}{\delta E^2-u^2|\delta{\bf p}|^2}
   = -\frac{1}{2} + \frac{\mathbb{P}}{2}\frac{\delta E^2}{\delta E^2-u^2|\delta{\bf p}|^2} \nonumber \ ,
\end{eqnarray}
where $\mathbb{P}$ stands for the principal part, and we defined $\delta{\bf p} = {\bf p}-{\bf p}'$ and $\delta E = (|{\bf p}|^2 - |{\bf p}'|^2)/2m$ for brevity. The symmetrized potential has two components. The first component $-1/2$ is momentum-independent, so it corresponds to a purely local attractive contact potential in real space. We have shown in the previous section that such a potential produces a bound-state. The second component is momentum-dependent and non-local in real space. It exists because we included finite momentum transfers ($\delta E \neq 0$). Since it has a repulsive character for some momenta, we must check if it can weaken or destroy the bound-state.

After symmetrization we are guarantied to get real eigenvalues $\epsilon$. However, the non-local Schrodinger equation is still too difficult to solve. We will now make our second ``self-consistent'' approximation, which will allow us to verify the survival of the bound-state in the presence of the non-local component of the potential. We will start with a hypothesis that the ground-state of two electrons is indeed a bound state, and use it to construct an effective purely local potential that approximates the realistic one. This approximation will overestimate the non-local component of the full potential, but accurately include its local component that by itself would support a bound state. Therefore, if this effective local potential has a bound state, so does the full one and the starting hypothesis is justified.

The assumed bound state has zero angular momentum, so its wavefunction amplitude is largest at the zero separation $|{\bf r}|=0$ between two electrons and gradually decreases as $|{\bf r}|$ grows. Consequently, replacing $\psi({\bf r}')$ in (\ref{SchrBS}) by $\psi({\bf r})$ would overestimate the effect of the non-local component of the full potential at $|{\bf r}|<a$. If the non-local part is repulsive, this overestimate would yield an upper bound for the binding energy $\epsilon<0$. On the other hand, if the non-local part is attractive at short distances, the overestimate would produce a deeper binding energy than the real one, not particularly useful for quantitative purposes. In either case, however, a bound-state is guarantied to exist if the overestimate supports it. Note that this procedure would clearly not modify the purely local part of the potential, which is short-ranged and attractive according to (\ref{Vpsym}).

We will actually self-consistently find a bound state only when electron interactions are mediated by phonons that disperse in all three dimensions. Purely two-dimensional phonons that we currently pursue produce a repulsive effective local potential at short distances, which we will derive first. Replacing $\psi({\bf r}')$ by $\psi({\bf r})$ in the symmetrized (\ref{SchrBS}) yields the following effective local potential
\begin{eqnarray}\label{U2D}
U({\bf r}) \!\!&=&\!\! \int \dd^2 r'\, V_{\textrm{s}}({\bf r},{\bf r}')
   = \int\frac{\dd^2 p}{(2\pi)^2}\,e^{i{\bf pr}}\,\widetilde{V}_{\textrm{s}}({\bf p},0) \\
\!\!&=&\!\! \int\frac{\dd^2 p}{(2\pi)^2}\,e^{i{\bf pr}}\, \left\lbrack
    -\frac{1}{2} + \frac{\mathbb{P}}{2}\frac{\left(\frac{|{\bf p}|^2}{2m}\right)^2}
                                            {\left(\frac{|{\bf p}|^2}{2m}\right)^2-u^2|{\bf p}|^2}
  \right\rbrack \nonumber \ ,
\end{eqnarray}
and a standard Schrodinger equation
\begin{equation}
-\frac{\boldsymbol{\nabla}^2}{m}\psi({\bf r})+g^2 U({\bf r})\psi({\bf r})=\epsilon\psi({\bf r}) \ .
\end{equation}
The local potential (\ref{U2D}) created by the two-dimensional phonons can be easily calculated:
\begin{eqnarray}\label{Inter2D}
U_{2\textrm{D}}({\bf r}) \!\!&=&\!\! \frac{1}{2} \int\frac{\dd^2 p}{(2\pi)^2}\,e^{i{\bf pr}}\,
   \mathbb{P} \frac{u^2|{\bf p}|^2}{\left(\frac{|{\bf p}|^2}{2m}\right)^2-u^2|{\bf p}|^2} \nonumber \\
 \!\!&=&\!\! \frac{1}{2}\int\limits_{0}^{\infty}\frac{\dd p}{2\pi}\,
   J_0(p|{\bf r}|)p\,\frac{\mathbb{P}}{\left(\frac{p}{2mu}\right)^2-1} \nonumber \\
 \!\!&=&\!\! -\frac{(mu)^2}{2}Y_0(2mu|{\bf r}|) \ ,
\end{eqnarray}
where $J_n(z)$ and $Y_n(z)$ are the Bessel functions of the first and second kind respectively. This potential is repulsive and logarithmically divergent at $|{\bf r}| \to 0$. It is also oscillatory at larger distances, alternating between positive and negative regions. A potential of this kind could support bound-states only if the typical electron momentum scales were of the order of $mu$. This is, however, not the case. The sound velocity $u$ is much smaller than the typical electron velocity.

The coupling of the quantum well electrons to the bulk phonons of the superconducting material is captured by the vertex (\ref{Ver3D}), and this gives rise to a slightly different Bethe-Salpeter equation (\ref{BS2}):
\begin{eqnarray}\label{BS2b}
&& \!\!\!\!\!\!\!\!\!\!\! F(p_\mu) = ig_0^2 \, G\left(\frac{P_\mu}{2}+p_\mu\right) G\left(\frac{P_\mu}{2}-p_{\mu}\right) \\
&& ~~~~ \times \int\frac{\dd^3 p_{\mu}'}{(2\pi)^3}\frac{\dd q_z^{\phantom{k}}}{2\pi}\,
      D({\bf p}-{\bf p}'+\hat{\bf z}q_z, \omega-\omega') F(p_\mu') \ . \nonumber
\end{eqnarray}
The electron-phonon coupling $g_0$ has different engineering dimensions than $g$ to match the changed dimensionality of the integral. We can immediately see that the procedure of deriving an effective Schrodinger equation can be translated directly to the case of three-dimensional phonons merely by changing the phonon dispersion from $u|{\bf p}-{\bf p}'|$ to $u|{\bf p}-{\bf p}'+\hat{\bf z}q_z|$ and integrating out $q_z$ in addition to ${\bf p}'$. Adapting (\ref{U2D}) in this manner yields:
\begin{eqnarray}\label{Inter3D}
U_{3\textrm{D}}({\bf r}) \!\!&=&\!\! \!\int\!\!\frac{\dd^2 p}{(2\pi)^2}\frac{\dd q_z}{2\pi}\,\frac{e^{i{\bf pr}}}{2} \left\lbrack
    \frac{\mathbb{P}\,\left(\frac{|{\bf p}|^2}{2m}\right)^2}
         {\left(\frac{|{\bf p}|^2}{2m}\right)^2-u^2|{\bf p}+\hat{\bf z}q_z|^2} -1 \right\rbrack \nonumber \\
\!\!&=&\!\! -\frac{\theta(a-|{\bf r}|)}{2\pi a^2 d} + U_{3\textrm{D}}'({\bf r}) \ ,
\end{eqnarray}
where
\begin{eqnarray}\label{BSpotcorr}
U_{3\textrm{D}}'({\bf r}) \!\!&=&\!\! \!\int\!\!\frac{\dd^2 p}{(2\pi)^2}\frac{\dd q_z}{2\pi}\,\frac{e^{i{\bf pr}}}{2}
    \frac{\mathbb{P}\,\left(\frac{|{\bf p}|^2}{2m}\right)^2}
         {\left(\frac{|{\bf p}|^2}{2m}\right)^2-u^2|{\bf p}|^2-u^2q_z^2} \nonumber \\
\!\!&=&\!\! -\frac{1}{4u^2}\int\limits_0^{2mu}\frac{\dd p}{2\pi}
    \left(\frac{p^2}{2m}\right)^2\frac{J_0(p|{\bf r}|)} {\sqrt{1-\left(\frac{p}{2mu}\right)^2}} \nonumber \\
\!\!&=&\!\! -\frac{3(mu)^3}{16}\!\!\!\!\phantom{F}_1\!F_2\left(\frac{5}{2};1,3;-(mu|{\bf r}|)^2\right) \ .
\end{eqnarray}
Therefore, the bulk phonons create the core attractive potential at $|{\bf r}|<a\sim\Lambda^{-1}$ that we analyzed before, and an additional oscillatory potential $U_{3\textrm{D}}'({\bf r})$ given by the generalized hypergeometric function $\!\!\!\!\phantom{F}_1\!F_2$ whose asymptotic behavior is:
\begin{equation}
\phantom{F}_1\!F_2\left(\frac{5}{2};1,3;-x^2\right) \approx \left\lbrace \begin{array}{ccc}
  1-\frac{5}{6}x^2+\frac{35}{192}x^4 & , & x<1 \\[0.05in]
  \frac{0.85}{x}\sin(2x) & , & x>10
\end{array} \right\rbrace \ .
\end{equation}
The extra potential $U_{3\textrm{D}}'$ is actually attractive at short distances, but recall that we have overestimated its contribution. At any rate, the overall scale $(mu)^3$ of this potential is small in comparison to that of the attractive core, which is determined by the momentum cut-offs. The most significant correction from the naive analysis in the previous section is that the amplitude of the attractive core potential is reduced by a half due to the phonon dynamics (recall that $\lambda \sim \lambda_0^{\phantom{2}} / d \equiv g_0^2 / d$). The Cooper pair bound-states survive and have the same binding energy by the order of magnitude that we calculated before.

So far we have treated the 2DES as being electronically isolated from its environment at all energy scales. However, in reality we hope to engineer the electron-phonon coupling in the 2DES by the proximity effect. This means that the 2DES is coupled to a bulk superconductor whose spectrum contains extended states at energies $|E-\mu| > \Delta_0$, where $\Delta_0$ is the superconductor's pairing gap. Only those electrons in the 2DES whose in-plane momentum is small enough to give them energy $|E-\mu| < \Delta_0$ can experience the true two-dimensional dynamics. Hence, we can patch our previous results by introducing a different momentum cutoff, $p < \Lambda_{\Delta_0} \ll \Lambda$, which is related to $\Delta_0$. Only the core attractive potential in (\ref{Inter3D}) depends on the cutoff, while such dependence of (\ref{Inter2D}) and (\ref{BSpotcorr}) can be neglected. The patched core potential is weaker and has a larger range $l_{\Delta_0}^{\phantom{1}} \sim \Lambda_{\Delta_0}^{-1} \gg a$ than the one written in (\ref{Inter3D}). This effect is due to the ability of electrons to tunnel across the superconductor-2DES interface and completely unrelated to the electron-phonon coupling $\lambda$. Thus, it will plague even strongly coupled superconductor interfaces. We will consider the naive formal $\lambda \to \infty$ limit to isolate this issue. The ensuing Cooper pair binding energy scale $E_0$ at $\lambda \to \infty$ must be much smaller than the naively obtained scale $1/ma^2$ in the section \ref{secBound}. Note that the binding energy reduction to $E_0$ has little to do with the increased range of the core potential in the weak-coupling limit. The reduction is mainly the consequence of the weaker attraction strength. Thus, we can see without any further calculation that $E_0$ cannot be (much) larger than $\Delta_0$ because this would not allow both the fermionic quasiparticle and Cooper pair two-dimensional bands to fit within the bulk superconductor's pairing gap. Some (small) enhancement of $E_0$ can perhaps occur due to the in-plane momentum conservation. The energy scale of retardation corrections is still much smaller than $E_0 \sim \Delta_0$.

With all this in mind, we conclude that the bulk longitudinal phonons of the superconducting material are significant mediators of the interactions among electrons in the quantum well. It is possible, though not necessary, that a branch of two-dimensional interface phonons exists in the heterostructure. Even though they create a logarithmically divergent repulsive potential between electrons at short distances, this potential is unlikely to jeopardize the existance of bound-state Cooper pairs due to its overall small scale and slow divergence. Note that this divergence is cut off below $|{\bf r}|<a$ and thus ultimately innocuous, just like the much more dramatic divergence of the Coulomb repulsion which we consider next.

\subsection{Coulomb interactions}\label{secCoulomb}

Our analysis has so far ignored the Coulomb interactions between electrons. Since they are the main adversary of the phonon-mediated pairing, we will investigate here whether they destroy the Cooper pair bound-state in two dimensions. The answer to this question is ultimately microscopic and system-dependent. However, we will argue that the condition for the bound-state survival in the 2DES is universally close to the condition for pairing in the superconducting material that creates the proximity effect, assuming that the 2DES thickness $d$ is sufficiently small. In other words, an insulating 2DES close to its superconducting transition will feature low energy bound-state Cooper pair excitations if the coupling between its electrons and the proximate superconductor's phonons is not weakened too much by the interface geometry.

Let us first analyze the condition for pairing in conventional BCS superconductors. The phonon-mediated interaction between electrons can be accurately captured by the Bethe-Salpeter equation with the phonon propagator in (\ref{Prop}). We may neglect phonon-mediated energy transfers between colliding electrons in weakly coupled superconductors ($\omega_{\textrm{D}} \ll E_{\textrm{f}}$), and approximate the phonon propagator $D(p_\mu)$ with the Fourier transform $U^{\textrm{p}}_{\textrm{BCS}}({\bf p})$ of the effective electron-electron interaction potential $U^{\textrm{p}}_{\textrm{BCS}}({\bf r})$ written in (\ref{Ueff}):
\begin{equation}
U^{\textrm{p}}_{\textrm{BCS}}({\bf p}) = -\lambda_0 \ .
\end{equation}
This expression is valid only below the cutoff momentum scale $\Lambda \sim a^{-1}$, where $a$ is the inter-atomic spacing. We can view $\lambda_0$ as being roughly constant at $p<\Lambda$ and zero at $p>\Lambda$. Interactions at larger momentum scales are negligible because they can be generated only by the electron coupling to higher phonon bands. Interacting electrons near the Fermi surface exchange momenta of the order of $p_{\textrm{f}}$ and smaller, which lie sufficiently below $\Lambda$ to allow treating $\lambda_0$ as a constant. The screened Coulomb interaction strength in momentum space is:
\begin{equation}\label{Coulomb1}
U^{\textrm{c}}_{\textrm{BCS}}({\bf p})=\frac{4\pi e^2}{p^2+p_{\textrm{s}}^2} \ ,
\end{equation}
where $l_{\textrm{s}}^{\phantom{1}} = p_{\textrm{s}}^{-1}$ is the screening length and we use the Gaussian units. Through $l_{\textrm{s}}^{\phantom{1}}$ we include the effect of all electrons on the net repulsive interaction between any two electrons. This interaction is not subject to any momentum cutoff since the corresponding real-space Coulomb potential $U^{\textrm{c}}_{\textrm{BCS}}(r) = e^2/r$ should hold at arbitrarily small $r \ll l_{\textrm{s}}$. We could in principle improve on this expression by incorporating the relativistic corrections and some elements of photon dynamics, but this is unnecessary because the typical electron velocities are sufficiently smaller than the speed of light.

The universal prerequisite for pairing is that the total electron-electron interaction be attractive at important momentum transfers which occur in most scattering events. This condition is
\begin{equation}\label{BCScond1}
U^{\textrm{p}}_{\textrm{BCS}}(p) + U^{\textrm{c}}_{\textrm{BCS}}(p) < 0 \quad,\quad p < p_{\textrm{f}} \ .
\end{equation}
in Fermi liquids. Let us take the superconducting Niobium as an example in order to estimate the relevant scales. In Niobium, $E_{\textrm{f}} = 5.32\textrm{ eV}$, $p_{\textrm{f}} = 1.24\times10^{-24}\textrm{ kg}\cdot\textrm{m}/\textrm{s}$ (the effective mass $m^*$ is approximately the same as the free electron mass $m$), $a=3.3 \textrm{\AA}$ and $\Lambda = 2\pi/a = 2.01\times10^{-24}\textrm{ kg}\cdot\textrm{m}/\textrm{s}$. We can estimate the screening length using the Thomas-Fermi formula \cite{Ashcroft1976}:
\begin{equation}
p_{\textrm{s}} = \sqrt{\frac{4}{\pi} m e^2 p_{\textrm{f}}} \ ,
\end{equation}
which yields $p_{\textrm{s}} = 1.77\times10^{-24}\textrm{ kg}\cdot\textrm{m}/\textrm{s}$. This is a typical low-temperature situation in metals and conventional superconductors based on them, $p_{\textrm{f}} \approx p_{\textrm{s}} \approx \Lambda$. Knowing this, we can write a simple approximate form of the condition (\ref{BCScond1}):
\begin{equation}\label{BCScond2}
\lambda_0 > \frac{4\pi e^2}{p_{\textrm{f}}^2} \ .
\end{equation}
The electron-phonon coupling must be sufficiently strong to give rise to superconductivity. Since the dynamics in BCS superconductors involves momentum exchanges near or below $p_{\textrm{f}}$, where (\ref{Coulomb1}) exhibits very little momentum dependence, we may use the effective coupling $\lambda_0' = \lambda_0 - 4\pi e^2 / p_{\textrm{f}}^2$ in BCS formulas such as (\ref{Gap3D}) to obtain the pairing gap, critical temperature, etc. This would include the net effect of phonon-mediated and Coulomb interactions on superconductivity.

Now let us obtain the analogous condition for the existance of bound-state Cooper pairs in an insulating quantum well that has an interface with a conventional superconductor. The phonon-mediated electron-electron interaction will be again represented by a static momentum-independent potential:
\begin{equation}
U_{\textrm{p}}({\bf p}) = -\lambda
\end{equation}
that holds at $p<\Lambda$. The effective coupling $\lambda$ between the 2DES electrons and the superconductor's phonons has different physical dimensions than the superconductor's $\lambda_0$ because $U_{\textrm{p}}({\bf p})$ is a two-dimensional Fourier transform of a short-range potential. A crude estimate $\lambda \lesssim \lambda_0 d^{-1}$ involving the 2DES thickness $d$ is based on the physical dimensions and the fact that the heterostructure geometry plays the main role in reducing the value of $\lambda$ below the one set by the bulk electron-phonon coupling in the superconductor. We are neglecting the retardation effects due to phonon dynamics, because we found in the previous section that they are small.

The effective Coulomb potential in momentum space has different behaviors at small and large momentum transfers. Small momentum transfers correspond to large distances between electrons. Since we are considering an insulating quantum well, its electrons cannot give rise to screening. However, the superconductor's electrons can partially screen the quantum well electrons by creating ``image charges'' at the interface. Each electron excitation in the quantum well is effectively turned into a dipole whose moment  points in the direction perpendicular to the well and has magnitude $\sim ed$, or $\sim e(d+l_{\textrm{B}})$ if the tunneling barrier is present (we will assume no tunneling barrier in the following discussion). The ensuing dipole-dipole interaction is repulsive:
\begin{equation}
U_{\textrm{c}}({\bf r})\sim\frac{e^2 d^2}{r^3}\quad,\quad r>p_{\textrm{s}}^{-1} \ .
\end{equation}
Note that the superconductor cannot efficiently provide ``image charges'' at electron separations $r$ smaller than its screening length $l_{\textrm{s}}^{\phantom{1}} = p_{\textrm{s}}^{-1}$. The two-dimensional Fourier transform of this potential is:
\begin{eqnarray}\label{DipolePotential}
U_{\textrm{c}}({\bf p}) \!\!&\sim&\!\! e^2 d^2 \int \dd^2 r\, \frac{e^{-i{\bf pr}}}{r^{3}}
  = 2\pi (ed)^2 \int\limits_{p_{\textrm{s}}^{-1}}^{\infty} \dd r\,\frac{J_0(pr)}{r^2} \nonumber \\
\!\!&=&\!\! 2\pi(ed)^2 \left\lbrack
  \!\!\!\!\phantom{F}_1 F_2 \left(-\frac{1}{2};\frac{1}{2},1;-\left(\frac{p}{2p_{\textrm{s}}}\right)^2\right) p_{\textrm{s}}
    - p \right\rbrack \nonumber \\
\!\!&\to&\!\! 2\pi(ed)^2 \left( p_{\textrm{s}} - p + \frac{p^2}{4p_{\textrm{s}}} \right) \quad,\quad p<p_{\textrm{s}} \ .
  \qquad\quad
\end{eqnarray}
The last line is obtained from the Taylor expansion of the generalized hypergeometric function $\!\!\!\!\!\phantom{F}_1F_2$ in the limit where this expression applies. At momentum transfers sufficiently above $p_{\textrm{s}}$ we must go back to the unscreened Coulomb potential
\begin{equation}
U_{\textrm{c}}({\bf r})\sim\frac{e^2}{r} \quad,\quad r<p_{\textrm{s}}^{-1}
\end{equation}
whose Fourier transform is:
\begin{eqnarray}
U_{\textrm{c}}({\bf p}) \!\!&\sim&\!\! \frac{2\pi e^2}{p_{\textrm{s}}} \times \!\!\!\!\!\phantom{F}_{1}F_{2}
        \left(\frac{1}{2};1,\frac{3}{2};-\left(\frac{p}{2p_{\textrm{s}}}\right)^2\right) \\
\!\!&=&\!\! \frac{2\pi e^2}{p} + \textrm{(small oscillations)} \quad,\quad p>p_{\textrm{s}} \nonumber \ .
\end{eqnarray}

Using the previously developed picture of the Cooper pair bound-state in the absence of Coulomb interactions, we would conclude that the main momentum transfers between the two electrons of a bound-state Cooper pair occur just below $p \sim a^{-1} \sim \Lambda$  in a relatively broad range of momenta (down to $\kappa \sim \xi^{-1}$). The effective electron repulsion is approximately given by (\ref{DipolePotential}) in this range of momentum transfers, and this conclusion is not altered by the reduced cutoff $\Lambda_{\Delta_0}$ due to the interface between the 2DES and the bulk superconductor. Even in the worst case, the Coulomb interaction effects can be weak enough for the bound-state to survive ($U_{\textrm{p}}+U_{\textrm{c}}<0$):
\begin{equation}\label{QWcond1}
\lambda > 2\pi e^2 d^2 p_{\textrm{s}} \ .
\end{equation}
We can compare this with (\ref{BCScond2}) by converting the 2D coupling $\lambda$ to the equivalent 3D coupling $\lambda d$ which is quantitatively comparable with the superconductor's electron-phonon coupling $\lambda_0$ or slightly smaller when $d$ is sufficiently small. We will also use the fact that $p_{\textrm{f}} \approx p_{\textrm{s}} \approx \Lambda$, and assume that the quantum well thickness $d$ is larger but not much larger than the typical inter-atomic separation $\sim p_{f}^{-1}$ of the superconductor. Then, we can rewrite (\ref{QWcond1}) as:
\begin{equation}\label{QWcond2}
\lambda d > \frac{2\pi e^2}{p_{\textrm{f}}^2}(p_{\textrm{f}}d)^3 \ .
\end{equation}
Since $\lambda d$ is of the order of $\lambda_0$, we see that the conditions (\ref{BCScond2}) and (\ref{QWcond2}) are not too far from each other. If the geometry of the coupling between the electrons in the quantum well and the superconductor's phonons does not make $\lambda$ too small, the existance of pairing in the superconductor can imply the existance of bound-state Cooper pairs in the quantum well despite the Coulomb repulsion.

It should be noted that the net interaction potential remains dominated by the Coulomb repulsion at extremely short distances between electrons. This might appear to pose a problem given that a bound state in two dimensions is guaranteed only when the potential is attractive at short distances. However, such short distances in our system can be probed only by scattering events whose momentum transfers lie above the cutoff. This in fact corresponds to electron scattering between different bands, which is suppressed at low temperatures. The Cooper pair bound-state involves low-energy band quasiparticles, and is shaped at energies and momenta that lie below the cutoff.

We can now apply the above analysis to make some extremely crude estimates of the critical temperature $T_{\textrm{c}}^{\textrm{2D}}$ for pairing and superconductivity in the 2DES, knowing the critical temperature $T_{\textrm{c}}$ of the proximate superconductor material. Assuming weak pairing in the superconductor, we can extract its \emph{effective} electron-phonon coupling from the formula (\ref{Gap3D}) using \cite{Abrikosov1975} $\Delta_0 \approx 1.76 T_{\textrm{c}}$:
\begin{equation}\label{F1}
\lambda_0' \sim \frac{2\pi^2}{m p_{\textrm{f}} \log\left(\frac{2\omega_{\textrm{D}}}{1.76 T_{\textrm{c}}}\right)} \ .
\end{equation}
We will treat the value obtained from this formula as empirical, which means that it incorporates both the true electron-phonon coupling $\lambda_0$ and the Coulomb repulsions from (\ref{BCScond2}):
\begin{equation}\label{F2}
\lambda_0' \sim \lambda_0 - \frac{4\pi e^2}{p_{\textrm{f}}^2} \ .
\end{equation}
The intrinsic proximity-generated coupling between the 2DES electrons and superconductor's phonons is of the order of $\lambda \sim \lambda_0/d$, but the effective coupling $\lambda'$ that determines the critical temperature $T_{\textrm{c}}^{\textrm{2D}}$ is also reduced by the Coulomb repulsions according to (\ref{QWcond2}):
\begin{equation}\label{F3}
\lambda' = \frac{\lambda_0}{d} - \frac{2\pi e^2}{p_{\textrm{f}}^2 d}(p_{\textrm{f}}d)^3 \ .
\end{equation}
Finally the critical temperature in the 2DES is related to the Cooper pair binding energy (\ref{BindingEnergy2}):
\begin{equation}\label{F4}
T_{\textrm{c}}^{\textrm{2D}} \sim |\epsilon| \sim E_0 \exp\left(-\frac{4\pi}{m\lambda'}\right) \ ,
\end{equation}
By combining the last four formulas we obtain:
\begin{eqnarray}\label{Tc2DES}
T_{\textrm{c}}^{\textrm{2D}} \!\!&\sim&\!\! E_0 \exp\left\lbrace -\frac{2p_{\textrm{f}}d}
    {\pi/\log\left(\frac{2\omega_{\textrm{D}}}{1.76 T_{\textrm{c}}}\right)
      +\frac{2me^2}{p_{\textrm{f}}}\left\lbrack 1-\frac{1}{2} (p_{\textrm{f}}d)^{3}\right\rbrack}\right\rbrace \nonumber \\
  \!\!&=&\!\! E_0 \exp\left\lbrace -\frac{2\frac{d}{d_{\textrm{max}}}}
    {m p_{\textrm{f}} e^2 d_{\textrm{max}}^2 \left\lbrack 1 - \left(\frac{d}{d_{\textrm{max}}}\right)^3 \right\rbrack}\right\rbrace
  \ .
\end{eqnarray}
Clearly, $T_{\textrm{c}}^{\textrm{2D}}$ is maximized in the $d\to 0$ limit, but will not exceed $T_{\textrm{c}}$ because $E_0 \sim \Delta_0$. There is also a critical 2DES thickness $d$ at which $T_{\textrm{c}}^{\textrm{2D}} \to 0$:
\begin{equation}
d_{\textrm{max}} \sim \frac{1}{p_\textrm{f}} \left\lbrack
    2 - \frac{\pi p_{\textrm{f}}}{me^2\log\left(\frac{2\omega_{\textrm{D}}}{1.76 T_{\textrm{c}}}\right)}
  \right\rbrack^{\frac{1}{3}} \ .
\end{equation}
If $d>d_{\textrm{max}}$, then the Coulomb interactions win in the 2DES and Cooper pairs cannot be formed.

The Niobium parameters $p_{\textrm{f}} = 1.24\times10^{-24}\textrm{ kg}\cdot\textrm{m}/\textrm{s}$, $T_{\textrm{c}} \approx 10 \textrm{ K} \approx 0.86 \textrm{ meV}$, $\omega_{\textrm{D}} \approx 275/2\pi \textrm{ K} \approx 3.77 \textrm{ meV}$ yield $d_{\textrm{max}} \sim 4.89 \textrm{ \AA}$. This may be too small, but is not too far from the more realistic values $d \sim 10 - 100 \textrm{ \AA}$. One must keep in mind that this estimate is extremely crude and does not by itself rule out Niobium as a material that can produce an observable proximity effect. An ab-initio microscopic calculation would be able to produce a more accurate estimate, but only experiments can really verify practicality. The present crude analysis is more useful for comparison purposes. Consider, for example, the best conventional superconductor MgB$_2$. Its relevant parameters \cite{Putti2003, Cappelluti2004, Pallecchi2006} are $T_{\textrm{c}} \approx 40 \textrm{ K} \approx 3.44\textrm{ meV}$, $E_{\textrm{f}} \sim 0.25 \textrm{ eV}$, $\omega_{\textrm{D}} \approx 1100/2\pi \textrm{ K} \sim 15.09\textrm{ meV}$. This material with multiple and anisotropic bands is not accurately described by the simple weak-coupling BCS theory, but we will nevertheless use the BCS relationships to make ballpark estimates. Taking the worst-case scenario $m^* \approx m$ gives the upper bound on the Fermi momentum $p_{\textrm{f}} \sim \sqrt {2m E_{\textrm{f}}} \approx 2.7\times10^{-25}\textrm{ kg}\cdot\textrm{m/s}$ and the worst conditions for superconductivity in the 2DES. Now, $d_{\textrm{max}} \sim 29.49 \textrm{ \AA}$. A quantum well of thickness $d = 20 \textrm{ \AA}$ would yield a critical temperature $T_{\textrm{c}}^{\textrm{2D}} \sim 0.83 E_0$, which is close to the superconductor's $T_c \sim 40 \textrm{ K}$ given that $E_0 \sim T_{\textrm{c}}$. In fact, the dependence $T_{\textrm{c}}^{\textrm{2D}}(d)$ is very weak until $d \sim d_{\textrm{max}}$ because the factor $1/m p_{\textrm{f}} e^2 d_{\textrm{max}}^2 \sim 0.09$ that appears in (\ref{Tc2DES}) is very small for MgB$_2$. Note that even if this estimate were off by a factor of 100, the ensuing pairing temperature scale $\sim 400 \textrm{ mK}$ would be detectable. Therefore, while all these approximations cannot be taken seriously for quantitative purposes, they indicate by orders of magnitude that the proximity-induced superconductivity via the phonon mechanism can be achieved as a matter of principle in realistic quantum well heterostructures.

\section{Cooper pair tunneling}\label{secTunneling}

Here we derive the effective action of a two-dimensional electron system (2DES) placed in contact with a bulk superconductor (SC) by focusing entirely on the virtual Cooper pair tunneling from the SC to the 2DES. We will not pay attention to phonons in this section, but their contribution to the effective action can be easily included via an extra two-body potential that we already discussed in detail. Our present goals are to find how the Cooper pair dynamics renormalizes the electron dynamics in the TI, and demonstrate that even the tunneling mechanism produces short-range attractive interactions which have a spin-independent component capable of forming triplet pairs in the 2DES.

\subsection{Singlet proximity effect}\label{secSinglet}

Many aspects of the interaction proximity effect can be derived from the continuum limit path integral in which we model the SC as a material with instantaneous short-range attractive interactions among electrons, and the 2DES as a multi-orbital band-insulator without any intrinsic interactions. We will use $\widetilde{t}$ to denote imaginary time in this entire section, and $\tau$ to label the 2DES electron's orbital degrees of freedom. The SC-2DES interface in the $xy$ plane is modeled simply by orbital-dependent electron hopping $t_\tau$ between the SC and 2DES surfaces. The simplest imaginary-time action of this kind is $S = S_\textrm{p} + S_\textrm{g}$:
\begin{eqnarray}\label{S1}
S_\textrm{p} \!\! &=& \!\! \int\limits_{z<0} \!\! \dd\widetilde{t} \; \dd^3 r \left\lbrack f_\sigma^\dagger \left(
      D_0 - \frac{D_i^2}{2m} - \mu_{\textrm{sc}} \right) f_\sigma^{\phantom{\dagger}}
      - U f_\uparrow^\dagger f_\downarrow^\dagger f_\downarrow^{\phantom{\dagger}} f_\uparrow^{\phantom{\dagger}}
      \right\rbrack  \nonumber \\
  && + \int\limits_{\textrm{2DES}} \!\! \dd\widetilde{t} \; \dd^2 r \left\lbrack \psi^\dagger \Biggl(
      D_0 + H_0 \right) \psi \nonumber \\
  && - \int\limits_{z<0} \!\! \dd z \sum_\tau t_\tau(z)
        \Bigl( \psi_{\tau\sigma}^\dagger \mathcal{I}_{{\bf r},{\bf r}+z\hat{\bf z}}^{\phantom{\dagger}}
        f_{\sigma}^{\phantom{\dagger}}(z) + h.c. \Bigr) \Biggr\rbrack \nonumber \\
S_\textrm{g} \!\! &=& \!\! \int \! \dd\widetilde{t} \, \dd^3 r \, \frac{1}{4} F^{\mu\nu} F_{\mu\nu} \ ,
\end{eqnarray}
where the SC's electron fields $f_\sigma$ are defined in the semi-infinite space $z<0$, while the TI's electron fields $\psi_{\tau\sigma}$ are two-dimensional with spin $\sigma$ and orbital $\tau$ degrees of freedom. The gauge field $A_\mu$ dynamics is governed by $S_\textrm{g}$ via the electromagnetic field tensor $F_{\mu\nu}$ in the entire space, while the coupling of matter to the gauge field is embedded in:
\begin{equation}\label{Igauge}
D_\mu = \partial_\mu - ieA_\mu \quad,\quad
\mathcal{I}_{12} = \exp \left( -ie \int\limits_{x_{\mu 1}}^{x_{\mu 2}} \dd x_\mu A_\mu \right) \ .
\end{equation}
Specifying the bare Hamiltonian $H_0$ of the 2DES's electrons is not essential for the present purposes. The SC and 2DES are coupled via the hopping term $t_\tau(z)$, whose magnitude depends on the hopping distance. The possible spatial separation of the 2DES's orbitals along the $z$-axis translates into the $\tau$ and $z$ dependence of hopping, while we neglect hopping processes in which the electron's $x$ and $y$ coordinates change (without missing any essential physics).

We wish to integrate out the SC's fields and obtain the effective action for the 2DES alone. To that end, we first apply the Hubbard-Stratonovich transformation and decouple the interaction in the SC using a singlet Cooper pair field $\phi$. Then, we use the Bogoliubov transformation to group together all terms quadratic in electron fields:
\begin{equation}
S_\textrm{p} \! = \!\! \int\limits_{z<0} \!\! \dd\widetilde{t} \; \dd^3 r \left\lbrack \frac{1}{U}|\phi|^2
  + \left( \begin{array}{cc} f_\uparrow^\dagger & f_\downarrow^{\phantom{\dagger}} \end{array} \right)
    \mathcal{H}_{\textrm{BdG}}
    \left( \begin{array}{c} f_\uparrow^{\phantom{\dagger}} \\ f_\downarrow^\dagger \end{array} \right)
  \right\rbrack + \int\limits_{\textrm{TI}}\cdots
\end{equation}
where
\begin{equation}\label{BdG1}
\mathcal{H}_{\textrm{BdG}} =
    \left( \begin{array}{cc} D_0 - \frac{D_i^2}{2m} - \mu_{\textrm{sc}} & \phi \\
    \phi^\dagger & D_0 + \frac{D_i^2}{2m} + \mu_{\textrm{sc}} \end{array} \right) \ .
\end{equation}
The Gaussian integration of $f_\sigma$ leads to:
\begin{eqnarray}\label{Seff1}
&& \!\!\!\!\!\! S_{\textrm{p}}^{\textrm{eff}} = \int\limits_{z<0} \!\! \dd\widetilde{t} \; \dd^3 r
     \left\lbrack \frac{1}{U}|\phi|^2
   - \textrm{tr}\log \mathcal{H}_{\textrm{BdG}} \right\rbrack \\
&& ~ + \int\limits_{\textrm{2DES}} \!\! \dd^2 r \dd\widetilde{t} \Biggl\lbrack \psi^\dagger \left(
      D_0 + H_0 \right) \psi \nonumber \\
&& ~ - \int\limits_{z<0} \!\! \dd z \sum_{\tau\tau'} t_\tau t_{\tau'} \left( \begin{array}{cc}
       \psi_{\tau\uparrow}^\dagger & \psi_{\tau\downarrow}^{\phantom{\dagger}} \end{array} \right)
    \mathcal{H}_{\textrm{BdG}}^{-1} \left( \begin{array}{c}
       \psi_{\tau'\uparrow}^{\phantom{\dagger}} \\ \psi_{\tau'\downarrow}^\dagger \end{array} \right)
  \Biggr\rbrack \ . \nonumber
\end{eqnarray}
We have redefined $\psi_{\tau\sigma}({\bf r},z) \to \psi_{\tau\sigma}({\bf r}) \mathcal{I}_{{\bf r},{\bf r}+z\hat{\bf z}}^{\phantom{\dagger}}$ in all three-dimensional integrals in order to maintain gauge invariance. The $\phi$ field implicitly contained in $\mathcal{H}_{\textrm{BdG}}$ is three-dimensional, but mediates interactions between the two-dimensional 2DES's electrons. We can invert the matrix $\mathcal{H}_{\textrm{BdG}}$ by representing it as:
\begin{eqnarray}
  \mathcal{H}_{\textrm{BdG}} &=& D_0 + T + \Phi \\
  T = -\left( \frac{D_i^2}{2m} + \mu_{\textrm{sc}} \right) \sigma^z ~ &,& ~
  \Phi = \textrm{Re}(\phi) \sigma^x - \textrm{Im}(\phi) \sigma^y \ , \nonumber
\end{eqnarray}
where $\sigma^a$ are Pauli matrices in the Nambu spinor space. The matrix $\Phi$ now contains all spatial variations brought about by the fluctuating field $\phi$, and we can express $\mathcal{H}_{\textrm{BdG}}^{-1}$ as a gradient expansion:
\begin{eqnarray}
\mathcal{H}_{\textrm{BdG}}^{-1} \!\! &=& \!\!
      \left( D_0 + T + \Phi \right)^{-1} \left( D_0 - T - \Phi \right)^{-1} \left( D_0 - T - \Phi \right) \nonumber \\
 \!\! &=& \!\!
      \left( D_0^2 - T^2 - \Phi^2 + \delta\Phi \right)^{-1} \left( D_0 - T - \Phi \right) \ , \\
 \!\! &=& \!\!
      \left( \sum_{n=0}^{\infty} \mathcal{D}^n \right) \left( D_0^2 - T^2 - \Phi^2 \right)^{-1} \left( D_0 - T - \Phi \right) \ ,
 \nonumber
\end{eqnarray}
where
\begin{eqnarray}
\mathcal{D} \!\! &=& \!\! -\left( D_0^2 - T^2 - \Phi^2 \right)^{-1} \delta\Phi \\[0.07in]
\delta\Phi \!\! &=& \!\! \left( D_0 - T - \Phi \right) \left( D_0 + T + \Phi \right) - \left( D_0^2 - T^2 - \Phi^2 \right)
  \nonumber \\ \!\! &=& \!\! \lbrack D_0 , T+\Phi \rbrack - \lbrace T , \Phi \rbrace \ . \nonumber
\end{eqnarray}
If $\phi = \textrm{const}$ then $\delta\Phi = 0$, so the operator $\mathcal{D}$ gathers only spatial and imaginary time (gauge-covariant) derivatives acting on $\phi$. Hence the effect of $\mathcal{D}$ can be neglected deep in the superconducting phase. Keeping only the $n=0$ term amounts to the mean-field approximation. We can also consider the effect of $D_0^2$ and kinetic energy $T^2$ small next to $\Phi^2 = |\phi|^2$ (dominated by the SC's pairing gap). If we substitute $\mathcal{H}_{\textrm{BdG}}^{-1}$ into (\ref{Seff1}), we reveal the generated pairing interactions in the 2DES at the lowest order of the gradient expansion:
\begin{eqnarray}
&& \!\!\!\!\!\! S_{\textrm{p}}^{\textrm{eff}} = \int\limits_{z<0} \!\! \dd\widetilde{t} \; \dd^3 r
     \left\lbrack \frac{1}{U}|\phi|^2
   - \textrm{tr}\log \mathcal{H}_{\textrm{BdG}} \right\rbrack \\
&& ~~~~ + \int\limits_{\textrm{2DES}} \!\! \dd\widetilde{t} \, \dd^2 r \Biggl\lbrace
          \psi^\dagger \left(D_0 + H_0 \right) \psi \nonumber \\
&& ~~~~ + \int\limits_{z<0} \!\! \dd z \sum_{\tau\tau'} \frac{t_\tau t_{\tau'}}{|\phi|^2}
     \biggl\lbrack - \Bigl(\phi \, \psi_{\tau\uparrow}^\dagger \psi_{\tau'\downarrow}^\dagger + h.c. \Bigr)  + \cdots
    \biggr\rbrack \Biggr\rbrace \ .\nonumber
\end{eqnarray}
The dots denote various terms involving the derivatives of $\phi$, and renormalizations of the bare 2DES Hamiltonian $H_0$.

In the last step we integrate out $\phi$ and thus reverse the original Hubbard-Stratonovich transformation. The dynamics of $\phi$ is shaped by:
\begin{eqnarray}\label{Sphi}
S_\phi \!\! &=& \!\! \int\limits_{z<0} \!\! \dd\widetilde{t} \; \dd^3 r \left\lbrack \frac{1}{U}|\phi|^2
         - \textrm{tr}\log \mathcal{H}_{\textrm{BdG}} \right\rbrack \\
&=& \!\! \int\limits_{z<0} \!\! \dd\widetilde{t} \; \dd^3 r \biggl\lbrack
           \phi^\dagger (\partial_0-2ieA_0)\phi + \frac{1}{2M}\bigl\vert(\partial_i-2ieA_i)\phi\bigr\vert^{2} \nonumber \\
&& \!\! ~~~~~~~~~~~~~ -t|\phi|^{2}+u|\phi|^{4} + \cdots \biggr\rbrack \nonumber
\end{eqnarray}
The general form of this action and the charge of $\phi$ are restricted by symmetries, but calculating the concrete values of the couplings such as $t$ and $u$ is not worth the needed effort. We will be satisfied by the knowledge that $t>0$ carves a potential minimum at a finite $|\phi|=|\phi_0|$ due to SC's superconductivity. The value of $|\phi_0|$ (in the units of energy used here) is given by the pairing gap in the SC material. Assuming that the SC material is not close to its phase transition to the normal state, $|\phi_0|$ is reasonably large and only small fluctuations $\delta\phi = \phi - \phi_0$ have a significant dynamical effect. Integrating out $\delta\phi$ produces the following effective action of the 2DES's electrons:
\begin{eqnarray}\label{Seff2}
&& \!\!\!\!\!\!\! S_{\textrm{p}}^{\textrm{eff}} = \int\limits_{\textrm{2DES}} \!\! \dd\widetilde{t} \, \dd^2 r
   \Biggl\lbrack \psi^\dagger \left(D_0 + H' \right) \psi \\
&& \!\!\!\! - \int\limits_{z<0} \!\! \dd z \sum_{\tau\tau'} \frac{t_\tau t_{\tau'}}{|\phi_0|^2}
     \Bigl(\phi_0 \, \psi_{\tau\uparrow}^\dagger \psi_{\tau'\downarrow}^\dagger + h.c. \Bigr) \nonumber \\
&& \!\!\!\! - \left( \int\limits_{z<0} \!\! \dd z \,
    \sum_{\lbrace \tau_i \rbrace} \frac{t_{\tau_1} t_{\tau_2} t_{\tau_3} t_{\tau_4}}{|\phi_0|^4} \right) U'
      \psi_{\tau_1\uparrow}^\dagger \psi_{\tau_2\downarrow}^\dagger
      \psi_{\tau_3\downarrow}^{\phantom{\dagger}} \psi_{\tau_4\uparrow}^{\phantom{\dagger}} + \cdots \Biggr\rbrack \nonumber
\end{eqnarray}
The operator $H'$ contains the renormalized orbital-mixing kinetic energy and the dots represent the effect of the neglected higher derivative terms and large-amplitude $\delta\phi$ fluctuations. The correction to $H_0$ in $H'$ is small compared to the intrinsic 2DES Hamiltonian $H_0$ when the SC's gap $|\phi_0| \sim \Delta_0$ is large.

The generated interaction in the 2DES depends on the renormalized interaction coupling $U'$, not the bare one $U$. The renormalization comes from (\ref{Sphi}) and pertains to the stiffness of order parameter fluctuations in the vicinity of $\phi_0$. We see that the generated interaction is attractive, but its strength decreases when the SC's gap $|\phi_0|$ grows. This may look counter-intuitive for a proximity effect, but is in fact natural having in mind the dynamical origin of the considered interaction. After all approximations, we have in fact obtained a perturbative result in which the SC-2DES hopping is a small energy scale in comparison to the SC's gap $|\phi_0|$. The interaction is induced by virtual tunneling of a Cooper pair from the SC to the 2DES and back, which involves four electron hoppings and corresponds to the fourth power of hopping constants $t_\tau$. The tunneling occurs within a thin layer of the SC near the interface, where $t_\tau(z)$ is not too small. Temporarily removing an electron from the SC costs energy $|\phi_0|$, so if the SC's superconductivity is robust its Cooper pairs find it hard to tunnel and mediate interactions in the 2DES. Recall, however, that interactions between the 2DES's electrons mediated by the nearby SC's phonons are still nearly as strong as inside the SC, and thus dominate when $\phi_0$ is large.

This derivation shows that electron tunneling from the SC generates attractive interactions in the 2DES that can give rise to both inter- and intra-orbital pairing in the absence of a symmetry to protect against orbital mixing. In addition to interactions, we obtained a direct order parameter coupling proportional to $\phi_0 \psi^\dagger \psi^\dagger + h.c.$, which is usually considered in the superconducting proximity effect literature. The latter explicitly breaks the global U(1) symmetry and produces pairing correlations in all circumstances, so we can say that the 2DES always hosts a singlet condensate as a result of the proximity effect. However, this effect is not equally significant in all circumstances. The actual magnitude of the order parameter $|\varphi_0|$ in the 2DES also depends on the spectrum of the renormalized Hamiltonian $H'$. If this spectrum contains gapless modes, as in the case of a helical Fermi liquid formed on the surface of a strong topological insulator, then $|\varphi_0|$ can be sizable as a result of the direct order parameter coupling. The attractive interaction would also immediately create a pairing instability in these circumstances. The net $|\varphi_0|$ can be calculated, for example, by applying the mean-field approximation to (\ref{Seff2}). On the other hand, if the electronic 2DES spectrum is gapped, such as in a topological insulator quantum well, then $|\varphi_0|$ can be very small. One way to see this is to construct from (\ref{Seff2}) an effective action of fully gapped Cooper pairs $\varphi \sim \psi_\uparrow \psi_\downarrow$ in the 2DES (which could exist as excitations below the single-electron gap):
\begin{eqnarray}
S_\varphi^{\textrm{eff}} \!\! &=& \!\!\!\!\!\! \int\limits_{\textrm{2DES}} \!\! \dd\widetilde{t} \, \dd^2 r
   \biggl\lbrack \varphi^\dagger (\partial_0-2ieA_0)\varphi
  +\frac{1}{2M}\bigl\vert(\partial_i-2ieA_i)\varphi\bigr\vert^{2} \nonumber \\
&& \!\! -a(\phi_0\varphi^\dagger+h.c.)+t'|\varphi|^{2}+u'|\varphi|^{4} + \cdots \biggr\rbrack \ .
\end{eqnarray}
The gap implies $t'>0$, so the saddle-point approximation estimates $|\varphi_0|\approx a|\phi_0|/2t'$. Recall that the parameter $a$ is proportional to the square of the SC-2DES tunneling constant, and inversely proportional to $|\phi_0|^2$. Therefore, robust superconductivity in the SC material, weak tunneling and a large quantum well bandgap all conspire to make $|\varphi_0|$ small. It cannot ever completely vanish, however. In order to make the quantum well's superconductivity $|\varphi_0|$ large, it is necessary to turn $t'$ negative. This can indeed be accomplished in a practical heterostructure device by exposing the quantum well to a biased gate across an insulating medium. An applied gate voltage $V_{\textrm{g}}$ can pull electrons from the SC into the 2DES, which is captured in the effective action by the increase of the 2DES's chemical potential $\mu \propto V_{\textrm{g}}$ hidden in $H_0$ and $H'$. Since Cooper pairs are charged, this chemical potential directly appears in $t' = t'_0 - 2\mu$. This is how one can tune a large $|\varphi_0|\approx \sqrt{|t'|/2u'}$. We have emphasized only this effect in our recent analysis of the topological insulator quantum wells placed in proximity to a superconductor \cite{Nikolic2011a, Nikolic2012a}.

In the last stage of the present derivation, we wish to affirm that the generated attractive interactions in the 2DES are short-ranged. For that purpose we must draw a distinction between the proximity effect created by a superconductor and a (hypothetical) superfluid. If the SC material were a superfluid, its matter field would not be coupled to the electromagnetic gauge field $A_\mu$. Then, the SC's Cooper pair action (\ref{Sphi}) with $A_\mu$ set to zero would contain the dynamics of gapless Goldstone modes, and integrating them out in the following step would generate an algebraic long-range (Coulomb-like) interaction among the 2DES's electrons. Goldstone modes are gapped out by the Anderson-Higgs mechanism in real superconductors. This can be easily seen by fixing a gauge to pass all local phase fluctuations of the matter fields such as $\phi$ onto the ``longitudinal'' modes of the gauge field $A_\mu$, via a gauge transformation:
\begin{equation}\label{GaugeTransf1}
  A_\mu \to A_\mu + \frac{1}{e}\partial_\mu \lambda \quad,\quad \phi \to \phi e^{2i\lambda} \quad\dots \ .
\end{equation}
The action of the fluctuations $\delta\phi = \phi-\phi_0$ derived from (\ref{Sphi}) acquires a term proportional to $e^2 |\phi_0|^2 (A_\mu A_\mu)$ in this gauge when the ground-state of the SC material is superconducting ($|\phi_0| \neq 0$), so all excitations in the SC, now described by $A_\mu$, are clearly gapped. Integrating out $\delta\phi$, therefore, produces short-range (Yukawa-like) attractive interactions in the 2DES.

The 2DES effective action (\ref{Seff2}) is still written as a gauge theory. The direct pairing term with $\phi_0 \psi_\uparrow^\dagger \psi_\downarrow^\dagger + h.c.$ looks very awkward when one recalls that the factors $\mathcal{I}_{12}$ from (\ref{Igauge}) are embedded in the electron operators $\psi_{\tau\sigma}({\bf r},z)$. We can get rid of those $\mathcal{I}_{12}$ factors only by integrating out $A_\mu$ and absorbing any background gauge field into $\phi_0$ by a gauge transformation. It is convenient to integrate out $A_\mu$ in the same gauge that we mentioned above in the context of the Anderson-Higgs mechanism. The background gauge field is then simply zero, and the direct order parameter coupling yields to the same form that is obtained in a superfluid proximity effect. Note that the global phase of $\phi_0$ cannot be eliminated by a gauge transformation, but its absolute physical meaning is dubious because it can be identified with an arbitrary choice of a reference phase at a single point of space-time (all other local phases are gauge-dependent). The physical meaning of the order parameter coupling is that the local phases in the 2DES are correlated with those of the SC material in any gauge.

Integrating out $A_\mu$ produces additional effect due to the fact that electromagnetic fields are not confined to the SC and 2DES volumes. Photons are gapless outside the system we consider and mediate the usual long-range Coulomb interactions between electrons. The Coulomb forces, however, are laterally screened out in the 2DES since its ground state exhibits phase correlations, and efficiently so when $|\varphi_0|$ is large as discussed earlier. Photon screening in the semi-infinite space occupied by the SC material also generates an image-charge effect for the 2DES's electrons. Every electron in the 2DES effectively becomes an electric dipole and interacts with other 2DES electrons accordingly at length-scales larger than the photon screening length.

\subsection{Triplet proximity effect}\label{secTriplet}

We have seen that the proximity to a superconductor (SC) creates via Cooper pair tunneling not only a direct order parameter coupling in the two-dimensional electron gas (2DES), but also an effective attractive interaction that qualitatively mirrors the interactions in the SC responsible for superconductivity. So far we have considered a particular spin-dependent part of this interaction that leads to the pairing in singlet channels. However, its phonon-mediated Coulomb origin introduces a spin-independent component to this interaction, which could create triplet Cooper pairs in principle if the Pauli exclusion could be overcome. Indeed, topological insulator quantum wells are 2DESs with orbital degrees of freedom that allow spin-triplet Cooper pairs to form as inter-orbital singlets, and even give them energetic advantage via the Rashba spin-orbit coupling. Here we will explore the ability of Cooper pair tunneling to generate the spin-independent interaction component in the 2DES.

The phonon-mediated attractive interactions in the SC do have a small finite range established by charge screening, which we will optimistically compare with the thickness $d$ of the quantum well 2DES. Taking this into account, we will find that the effective attraction inside the quantum well could also bind spinful inter-orbital triplet pairs, and we will estimate its strength. But, we need a more complicated model in order to capture the inter-orbital pairing in the quantum well. Consider the following action for the SC material:
\begin{eqnarray}\label{S2}
S_{\textrm{sc}} \!\! &=& \!\! \int\limits_{z<0} \!\! \dd\widetilde{t} \, \dd^3 r \Biggl\lbrack f_\sigma^\dagger \left(
      D_0 - \frac{D_i^2}{2m} - \mu_{\textrm{sc}} \right) f_\sigma^{\phantom{\dagger}}
  \\ && - \frac{1}{2} \int\limits_{z<0} \!\! \dd^3 r' \, V({\bf r}-{\bf r}')
        f_\sigma^\dagger({\bf r}) f_{\sigma}^{\phantom{\dagger}}({\bf r})
        f_{\sigma'}^\dagger({\bf r}') f_{\sigma'}^{\phantom{\dagger}}({\bf r}')
      \Biggr\rbrack \nonumber
\end{eqnarray}
in which we neglect the spin dependence of the attractive interaction $V>0$, assuming its origin in the Coulomb-based phonon mechanism. Decoupling it by the Cooper channel Hubbard-Stratonovich transformation requires complex fields $\phi_{\sigma\sigma'}({\bf r},{\bf r}')$ that have a doubled position and spin dependence. In order to avoid ambiguity, one could temporarily reorganize the above conventional double integral over coordinates in a manner that samples every pair $({\bf r},{\bf r}')$ only once, rather than twice (with opposite ordering, which requires the corrective factor of $\frac{1}{2}$). Each interaction term, defined by a pair $({\bf r},{\bf r}')$ and one of the four combinations $(\sigma,\sigma')$, can then be decoupled by a unique complex field $\phi_{\sigma\sigma'}({\bf r},{\bf r}')$. But in order to write the result using the conventional double integral, we must create a copy of the Hubbard-Stratonovich field via $\phi_{\sigma\sigma'}({\bf r},{\bf r}') = -\phi_{\sigma'\sigma}({\bf r}',{\bf r})$:
\begin{eqnarray}\label{S2a}
&& S_{\textrm{sc}} = \int\limits_{z<0} \!\! \dd\widetilde{t} \, \dd^3 r \, f_\sigma^\dagger \left(
      D_0 - \frac{D_i^2}{2m} - \mu_{\textrm{sc}} \right) f_\sigma^{\phantom{\dagger}} \\
&& ~~~~ + \frac{1}{2} \!\!\! \int\limits_{z,z'<0} \!\!\! \dd\widetilde{t} \; \dd^3 r \, \dd^3 r'
     \, \sum_{\sigma\sigma'} \Biggl\lbrack \frac{1}{V({\bf r}-{\bf r}')} |\phi_{\sigma\sigma'}({\bf r},{\bf r}')|^2 \nonumber \\
&& ~~~~~~~~~~~ + \phi_{\sigma\sigma'}^{\phantom{\dagger}}({\bf r},{\bf r}') f_\sigma^\dagger({\bf r}) f_{\sigma'}^\dagger({\bf r}')
   + h.c. \Biggr\rbrack \ . \nonumber
\end{eqnarray}
Note that $V({\bf r}-{\bf r}') \to 0$ at large $|{\bf r}-{\bf r}'|$ merely quenches the fluctuations of the corresponding $\phi_{\sigma\sigma'}({\bf r},{\bf r}')$ and prevents them from mediating long-range forces. This is also what keeps the action scale linearly with the system volume. The gauge transformations of $\phi_{\sigma\sigma'}({\bf r},{\bf r}')$ are:
\begin{equation}\label{GaugeTransf2}
  \phi_{\sigma\sigma'}({\bf r},{\bf r}') \to \phi_{\sigma\sigma'}({\bf r},{\bf r}')
     e^{i\lambda({\bf r})} e^{i\lambda({\bf r}')} \ .
\end{equation}

Next, we need to express the action in the Bogoliubov-de Gennes form in order to integrate out $f_\sigma$. This time, however, the presence of $\phi_{\sigma\sigma'}({\bf r},{\bf r}')$ makes the Bogoliubov-de Gennes Hamiltonian $\mathcal{H}_{\textrm{BdG}}$ highly non-diagonal in the coordinate representation. Let us organize all SC's fields into a vector $F$ whose four-component blocks $F({\bf r}) = \left( f_\uparrow^{\phantom{\dagger}}({\bf r}), f_\downarrow^\dagger({\bf r}), f_\downarrow^{\phantom{\dagger}}({\bf r}), f_\uparrow^\dagger({\bf r}) \right)$ are indexed by the coordinates ${\bf r}$. Similarly, $\mathcal{H}_{\textrm{BdG}}$ is a block-matrix whose blocks are indexed by coordinate pairs $({\bf r},{\bf r}')$:
\begin{widetext}
\begin{equation}
\sst \mathcal{H}_{\textrm{BdG}}({\bf r},{\bf r}') ~=~ \left( \begin{array}{cccc}
    \sst \left(D_0 - \frac{D_i^2}{2m} - \mu_{\textrm{sc}}\right) \delta({\bf r}'-{\bf r}) &
    \sst \phi_{\uparrow\downarrow}^{\phantom{*}}({\bf r},{\bf r}') &
    \sst 0 &
    \sst \phi_{\uparrow\uparrow}^{\phantom{*}}({\bf r},{\bf r}')
  \\
    \sst \phi_{\uparrow\downarrow}^*({\bf r}',{\bf r}) &
    \sst \left(D_0 + \frac{D_i^2}{2m} + \mu_{\textrm{sc}}\right) \delta({\bf r}'-{\bf r}) &
    \sst \phi_{\downarrow\downarrow}^*({\bf r}',{\bf r}) &
    \sst 0
  \\
    \sst 0 &
    \sst \phi_{\downarrow\downarrow}^{\phantom{*}}({\bf r},{\bf r}') &
    \sst \left(D_0 - \frac{D_i^2}{2m} - \mu_{\textrm{sc}}\right) \delta({\bf r}'-{\bf r}) &
    \sst \phi_{\downarrow\uparrow}^{\phantom{*}}({\bf r},{\bf r}')
  \\
    \sst \phi_{\uparrow\uparrow}^*({\bf r}',{\bf r}) &
    \sst 0 &
    \sst \phi_{\downarrow\uparrow}^*({\bf r}',{\bf r}) &
    \sst \left(D_0 + \frac{D_i^2}{2m} + \mu_{\textrm{sc}}\right) \delta({\bf r}'-{\bf r})
\end{array} \right) \ .
\end{equation}
\end{widetext}
In order to represent the SC-2DES interface tunneling from ~(\ref{S1}), we must also reorganize the 2DES's fields $\psi_{\tau\sigma}$ into a vector $\Psi$ compatible with $F$. $\Psi$ must be made of four-component blocks indexed by the coordinates ${\bf r}$, and cannot have any leftover orbital dependence. These restrictions lead us to the blocks:
\begin{equation}
\Psi({\bf r}) \!=\! \sum_\tau \! t_\tau(z) \! \left(
  \psi_{\tau\uparrow}^{\phantom{\dagger}} (x,y), \psi_{\tau\downarrow}^\dagger (x,y),
  \psi_{\tau\downarrow}^{\phantom{\dagger}} (x,y), \psi_{\tau\uparrow}^\dagger (x,y)
\right)
\end{equation}
The Bogoliubov-de Gennes form of the complete action is:
\begin{eqnarray}\label{S2b}
S \!\! &=& \!\!\!\!\! \int\limits_{\textrm{2DES}} \!\! \dd\widetilde{t} \, \dd^2 r \,
     \psi^\dagger \left(D_0 + H_0 \right) \psi \\
&& + \frac{1}{2} \!\!\! \int\limits_{z,z'<0} \!\!\! \dd\widetilde{t} \; \dd^3 r \, \dd^3 r' \,
     \sum_{\sigma\sigma'} \frac{|\phi_{\sigma\sigma'}({\bf r},{\bf r}')|^2}{V({\bf r}-{\bf r}')} \nonumber \\
&& + \frac{1}{2} F^\dagger \mathcal{H}_{\textrm{BdG}} F
   - \frac{1}{2} \left( \Psi^\dagger \mathcal{I} F + h.c.\right) \ . \nonumber
\end{eqnarray}
The matrix $\mathcal{I}$ in the tunneling term is required by gauge invariance, and its elements are given by (\ref{Igauge}) and
\begin{equation}
\mathcal{I}({\bf r},{\bf r}') =
  \textrm{diag} (\mathcal{I}_{x\hat{\bf x}+y\hat{\bf y},{\bf r}'}^{\phantom{\dagger}},
                 \mathcal{I}_{x\hat{\bf x}+y\hat{\bf y},{\bf r}'}^\dagger,
                 \mathcal{I}_{x\hat{\bf x}+y\hat{\bf y},{\bf r}'}^{\phantom{\dagger}},
                 \mathcal{I}_{x\hat{\bf x}+y\hat{\bf y},{\bf r}'}^\dagger) \nonumber \ .
\end{equation}
Integrating out the SC's fields amounts to integrating out the Grassmann number components of the vector $F$. Since $F$ groups together all available SC's Grassmann fields, and not just a half of them as usual, this Gaussian integration produces the Pfaffian of the matrix $\mathcal{H}_{\textrm{BdG}}$, which we immediately relate to the determinant, $\textrm{Pf}(\mathcal{H}_{\textrm{BdG}}) = \sqrt{\textrm{det}(\mathcal{H}_{\textrm{BdG}})}$. The resulting effective theory of electrons inside the TI is:
\begin{eqnarray}\label{S2c}
S_{\textrm{eff}} \!\! &=& \!\!\!\!\! \int\limits_{\textrm{2DES}} \!\! \dd\widetilde{t} \, \dd^2 r \,
      \psi^\dagger \left(D_0 + H_{\textrm{g}} \right) \psi \\
&& + \frac{1}{2} \!\!\! \int\limits_{z,z'<0} \!\!\! \dd\widetilde{t} \; \dd^3 r \, \dd^3 r' \,
     \sum_{\sigma\sigma'} \frac{|\phi_{\sigma\sigma'}({\bf r},{\bf r}')|^2}{V({\bf r}-{\bf r}')} \nonumber \\
&& - \frac{1}{2} \Psi^\dagger \mathcal{I} \mathcal{H}_{\textrm{BdG}}^{-1} \mathcal{I}^\dagger \Psi
   - \frac{1}{2} \textrm{tr}\log \mathcal{H}_{\textrm{BdG}} \ . \nonumber
\end{eqnarray}

In order to learn something about the inter-orbital triplet pairing we need to analyze the inverse Bogoliubov-de Gennes matrix $\mathcal{H}_{\textrm{BdG}}^{-1}$. To that end, let us separate $\mathcal{H}_{\textrm{BdG}} = \mathcal{H}_{\textrm{s}} + \mathcal{H}_{\textrm{t}}$ into pieces that couple electrons of opposite and equal spins, $\mathcal{H}_{\textrm{s}}$ and $\mathcal{H}_{\textrm{t}}$ respectively. Using the block-notation,
\begin{eqnarray}
\mathcal{H}_{\textrm{t}}({\bf r},{\bf r}') &=& \left( \begin{array}{cccc}
    \sst 0 &
    \sst 0 &
    \sst 0 &
    \sst \phi_{\uparrow\uparrow}^{\phantom{*}}({\bf r},{\bf r}')
  \\
    \sst 0 &
    \sst 0 &
    \sst \phi_{\downarrow\downarrow}^*({\bf r}',{\bf r}) &
    \sst 0
  \\
    \sst 0 &
    \sst \phi_{\downarrow\downarrow}^{\phantom{*}}({\bf r},{\bf r}') &
    \sst 0 &
    \sst 0
  \\
    \sst \phi_{\uparrow\uparrow}^*({\bf r}',{\bf r}) &
    \sst 0 &
    \sst 0 &
    \sst 0
\end{array} \right) \nonumber
 \\[0.1in]
\mathcal{H}_{\textrm{s}} &=& \mathcal{H}_{\textrm{BdG}} \Bigr\vert_{\phi_{\uparrow\uparrow} = \phi_{\downarrow\downarrow} = 0}
  \ .
\end{eqnarray}
The $\mathcal{H}_{\textrm{s}}$ part contains various singlet fields $\phi_{\sigma\bar{\sigma}}$, where $\bar{\sigma}=-\sigma$. Since $\phi_{\sigma\bar{\sigma}}$ at two different locations $({\bf r}_1,{\bf r}_1')$ and $({\bf r}_2,{\bf r}_2')$ are not isolated by any physical symmetry, the effective action contains phase-locking couplings $\phi_{\sigma\bar{\sigma}}^\dagger ({\bf r}_1,{\bf r}_1') \phi_{\sigma\bar{\sigma}}^{\phantom{\dagger}} ({\bf r}_2,{\bf r}_2') + h.c.$ that make all fields $\phi_{\sigma\bar{\sigma}}$ simultaneously condensed in the SC. The condensate amplitudes are regulated by the microscopic interaction $V({\bf r}-{\bf r}')$ and decay rapidly when the separation between the end-points ${\bf r}$ and ${\bf r}'$ grows. At any rate, $\mathcal{H}_{\textrm{s}}$ is characterized by a finite energy scale $|\phi_0|$ that corresponds to the superconducting order parameter. In contrast, the triplet fields $\phi_{\sigma\sigma}$ are not condensed in the SC. Their fluctuations at distances comparable to the quantum well thickness, $|{\bf r}-{\bf r}'| \approx d$, have an effect on the triplet pairing in the TI, but the amplitude of these fluctuations can be considered small in comparison to $|\phi_0|$. Therefore, we may treat $\mathcal{H}_{\textrm{t}}$ as a perturbation to $\mathcal{H}_{\textrm{s}}$ and write:
\begin{eqnarray}
&& \mathcal{H}_{\textrm{BdG}}^{-1} = (\mathcal{H}_{\textrm{s}}^{\phantom{1}} + \mathcal{H}_{\textrm{t}}^{\phantom{1}})^{-1} =
   \mathcal{H}_{\textrm{s}}^{-1} \sum_{n=0}^{\infty}
     \left( - \mathcal{H}_{\textrm{t}}^{\phantom{1}} \mathcal{H}_{\textrm{s}}^{-1} \right)^n \nonumber \\
&& ~~~~~~~~ \approx \mathcal{H}_{\textrm{s}}^{-1}
    \left( \mathcal{H}_{\textrm{s}}^{\phantom{1}} - \mathcal{H}_{\textrm{t}}^{\phantom{1}} \right)
    \mathcal{H}_{\textrm{s}}^{-1} \ .
\end{eqnarray}
The generated quadratic coupling of $\Psi$ in (\ref{S2c}) now takes the form $\widetilde{\Psi}^\dagger \left( \mathcal{H}_{\textrm{s}} - \mathcal{H}_{\textrm{t}} \right) \widetilde{\Psi}$, where $\widetilde{\Psi} = \mathcal{H}_{\textrm{s}}^{-1} \mathcal{I}^\dagger \Psi$. Since $\mathcal{H}_{\textrm{s}}$ contains no spinful triplets $\phi_{\sigma\sigma}$, the components of $\widetilde{\Psi}$ are operators that create or annihilate excitations with well-defined spin $S^z=\pm\frac{1}{2}$, despite being linear combinations of bare electron creation and annihilation operators. Hence, we can think of $\widetilde{\Psi}$ as the vector of quasiparticle field operators, which can further interact via the $\widetilde{\Psi}^\dagger \left( \mathcal{H}_{\textrm{s}} - \mathcal{H}_{\textrm{t}} \right) \widetilde{\Psi}$ term. Specifically, we are interested in their triplet-forming interactions:
\begin{eqnarray}
&& \!\!\!\!\!\!\!\! \frac{1}{2} \widetilde{\Psi}^\dagger \mathcal{H}_{\textrm{t}} \widetilde{\Psi} \approx
  \frac{1}{2} \!\!\! \int\limits_{z,z'<0} \!\!\! \dd\widetilde{t} \; \dd^3 r \, \dd^3 r' \, \sum_{\tau\tau'}
  \frac{t_{\tau}(z) t_{\tau'}(z')}{|\phi_0|^2} \\
&& \! \times \sum_\sigma
  \left\lbrack \phi_{\sigma\sigma}^{\phantom{\dagger}}({\bf r},{\bf r}')
      \mathcal{I}_{{\bf r}_\perp^{\phantom{k}},{\bf r}}^{\phantom{\dagger}}
      \mathcal{I}_{{\bf r}_\perp',{\bf r}'}^{\phantom{\dagger}}
    \widetilde{\psi}_{\tau\sigma}^\dagger({\bf r}_\perp^{\phantom{k}}) \widetilde{\psi}_{\tau'\sigma}^\dagger({\bf r}_\perp')
  + h.c. \right\rbrack \ , \nonumber
\end{eqnarray}
where ${\bf r}_\perp=(x,y)$ and we used the individual components $\widetilde{\psi}_{\tau\sigma}({\bf r}_\perp)$ of $\widetilde{\Psi}$ defined by:
\begin{equation}
\widetilde{\Psi}({\bf r}) = \sum_\tau \frac{t_\tau(z)}{|\phi_0|} \!
   \left(
     \mathcal{I}_{{\bf r}_\perp,{\bf r}}^\dagger \widetilde{\psi}_{\tau\uparrow}^{\phantom{\dagger}} ,
     \mathcal{I}_{{\bf r}_\perp,{\bf r}}^{\phantom{\dagger}} \widetilde{\psi}_{\tau\downarrow}^\dagger ,
     \mathcal{I}_{{\bf r}_\perp,{\bf r}}^\dagger \widetilde{\psi}_{\tau\downarrow}^{\phantom{\dagger}} ,
     \mathcal{I}_{{\bf r}_\perp,{\bf r}}^{\phantom{\dagger}} \widetilde{\psi}_{\tau\uparrow}^\dagger
   \right) \nonumber
\end{equation}
This definition ensures that the quasiparticle operators $\widetilde{\psi}_{\tau\sigma}$ have the same dimensions, quantum numbers and normalization as the 2DES's electron operators $\psi_{\tau\sigma}$. The dynamics of uncondensed triplets $\phi_{\sigma\sigma}$ can be approximated by a Gaussian:
\begin{equation}
S_{\textrm{eff}}(\phi_{\sigma\sigma}) \approx \frac{1}{2} \!\!\! \int\limits_{z,z'<0} \!\!\! \dd^3 r \dd^3 r' \dd\widetilde{t}
     \; \sum_{\sigma} \frac{|\phi_{\sigma\sigma}({\bf r},{\bf r}')|^2}{V'({\bf r}-{\bf r}')} + \cdots
\end{equation}
with a renormalized positive coupling $V'$. It is then straight-forward to integrate out $\phi_{\sigma\sigma}$ and obtain the induced interactions in the triplet channels:
\begin{eqnarray}
&& S_{\textrm{eff}}^{\textrm{tri}} = -\frac{1}{2} \!\!\! \int\limits_{z,z'<0} \!\!\! \dd\widetilde{t} \; \dd^3 r \, \dd^3 r' \,
  V'({\bf r}-{\bf r}') \sum_\sigma \\
&& ~~~~~~ \times
    \sum_{\lbrace \tau_i \rbrace} \frac{t_{\tau_1}(z) t_{\tau_2}(z') t_{\tau_3}(z') t_{\tau_4}(z)}{|\phi_0|^4} \nonumber \\
&& ~~~~~~ \times
      \widetilde{\psi}_{\tau_1\sigma}^\dagger({\bf r}_\perp^{\phantom{k}})
      \widetilde{\psi}_{\tau_2\sigma}^\dagger({\bf r}_\perp')
      \widetilde{\psi}_{\tau_3\sigma}^{\phantom{\dagger}}({\bf r}_\perp')
      \widetilde{\psi}_{\tau_4\sigma}^{\phantom{\dagger}}({\bf r}_\perp^{\phantom{k}})
  + \cdots \ . \nonumber
\end{eqnarray}
Note that the effective action in the 2DES does not contain a direct order parameter coupling in the triplet channel, because the SC has no triplet condensation. A direct coupling between 2DES's triplet pairs and SC's singlet pairs is forbidden by symmetry.

We conclude that the Cooper pair tunneling mediates effective interactions among the TI's electrons that are short-ranged, attractive and weak if the SC's pairing gap $|\phi_0|$ is large in comparison to the electron's SC-TI hopping energy scales $t_\tau(z\approx 0)$. The induced pairing tendency at spatial separations $r$ is related to the actual phonon-mediated attractive interaction $V(r)$ that forms Cooper pairs in the SC. The induced interaction in the TI is generally capable of mixing the orbital content and forming inter-orbital triplets as well as intra-orbital singlets.

We must note that the strength of inter-orbital pairing in the quantum well 2DES, mediated by the virtual Cooper pair tunneling, depends on the renormalized SC's bulk interaction potential at electron separations comparable to the quantum well thickness, $V'(d)$. This may be much smaller than the direct intra-orbital singlet pairing potential for two reasons. First of all, the quantum well may be a few crystal unit-cells thick, so that $d$ is somewhat larger than the bulk SC's screening length $l_{\textrm{s}}$, which is roughly equal to the inter-atomic separation. Even though $d \sim l_{\textrm{s}}$ by order of magnitude, the exponential suppression of the screened interaction strength with distance can make $V'(d)$ weak. Second, the interaction renormalization in the SC's bulk is contributed by the statistical (Pauli) repulsion between electrons of the same spin within the SC. This surely weakens $V'(r)$ at separations $r\sim d$ where electrons are quite delocalized. For these reasons, triplet pairing in the quantum well is most likely shaped by the direct SC's phonon mechanism, whose backbone Coulomb forces are spin-independent and felt across the quantum-well thickness without being jeopardized by screening, as we emphasized earlier. In contrast, singlet pairing is contributed by the Cooper pair tunneling, but not strongly unless the hopping $t_\tau$ is unusually large or the SC order parameter $|\phi_0|$ is small.

\section{Discussion and conclusions}

In this paper we qualitatively analyzed the influence of a conventional superconductor on the electron dynamics in another non-superconducting material near their common interface. We identified two mechanisms of influence, the coupling between the material's electrons and superconductor's phonons, and the virtual Cooper pair tunneling across the interface. We found that this yields induced effective interactions among the material's electrons near the interface, as well as a direct symmetry-breaking term in their effective action that constitutes the conventional proximity effect.

The phonon proximity effect creates a retarded spin-independent attractive interaction that can in principle be large enough to overcome the Coulomb repulsion and lead to independent Cooper pairing among the non-superconducting material's surface electrons. We calculated the minimum electron-phonon coupling across the interface that leads to this proximity-induced pairing. Its value is related to the superconductor's electron-phonon coupling, the spatial extent $d$ of the material's affected electrons near the surface, and the crystal's elasticity near the interface. Our generic conclusion is that any heterostructure with a sufficiently small $d$ and thin interface will satisfy the pairing condition, by the virtue of the analogous condition being satisfied inside the bulk superconductor.

If the superconductor's phonon-mediated pairing indeed occurs among the material's surface electrons, then the critical temperature of the resulting two-dimensional superconductor is mostly independent of its surface density. This does not hold at very low densities, near the surface superconductor-insulator transition. However, the character of the surface superconductivity changes. Cooper pairs can exist as two-body bound states in two dimensions, so their quantum phase transition is driven by quantum fluctuations, not pairing. These bound-state pairs are as large as the pairs in a dense two-dimensional BCS superconductor, so the effect of such fluctuations is visible only at extremely low densities where they do not overlap. More importantly, bound-state Cooper pairs can exist as gapped bosonic excitations in a (band) insulating state of the material's surface electrons, whose energy lies within the fermionic bandgap. Their binding energy is of the same order as the critical temperature of a dense BCS surface superconductor, and their gap can be controlled by a gate voltage. The existance and accessibility of the ``fluctuation regime'' dominated by the dynamics of surface bound Cooper pairs is perhaps the most interesting perk of the interaction proximity effect because it can enable engineering novel correlated two-dimensional states of quantum matter.

The tunneling mechanism, on the other hand, imprints the superconductor's electron-electron interactions on the material's surface electrons via the superconductor's Cooper pair dynamics. The induced interaction here is of the same kind as in the superconductor, i.e. short-ranged, attractive, and acting among electrons of any spin state. However, its strength is spin-dependent because the tunneling Cooper pairs are singlets. The induced surface interaction in the triplet channel is expected to be weak, but it may nevertheless be helpful to the phonon-mediated spin-independent interaction in stabilizing triplet Cooper pairs against the Coulomb repulsion. This is particularly relevant for materials that provide orbital electronic degrees of freedom, such as topological insulator quantum wells. They can host orbital-singlet spin-triplet Cooper pairs whose energy is either not prohibitively high, or even very competitive as a result of the spin-orbit coupling. Correlated quantum phases of triplet Cooper pairs in topological insulator quantum wells are a particularly attractive possibility because quantum fluctuations can in principle stabilize novel incompressible quantum liquids with non-Abelian quasiparticles.

The Cooper pair tunneling is most effective in the surface singlet channels, which are directly coupled to the superconductor's order parameter. The effective action contains an explicit U(1) symmetry-breaking term as a result of this, the usually considered proximity effect. Such a term always produces singlet pairing correlations among the material's surface electrons, but their magnitude also depends on the bare surface electron's spectrum. If the material's surface is normally metallic, the ensuing pairing correlations will be strong. This effect is particularly interesting in the topologically protected helical metals on the surfaces of bulk topological insulators. The naively singlet pairing term actually gives rise to a very interesting surface superconducting state shaped by the spin-orbit coupling, whose topological defects bind Majorana quasiparticles. The analogous effect on band-insulating surfaces, i.e. topological insulator quantum wells, is diminished or absent. There, the pairing term is a small perturbation to the surface bandgap, so the actual intrinsic singlet correlations are small. Instead, the main effect of Cooper pair tunneling are induced interactions in the singlet channels. They can shape bound-state Cooper pairs below the fermionic bandgap, whose dynamics can be made important by manipulating the surface chemical potential via the gate voltage. The ensuing singlet pairing is normally poised to win over triplets, unless the triplet dynamics becomes enhanced by the spin-orbit coupling.

The main motivation for this and our previous work\cite{Nikolic2011a, Nikolic2012a} is the prospect of creating fractional incompressible quantum liquids in topological insulator quantum wells. The emerging picture is that there are no fundamental obstacles to obtaining such novel strongly correlated quantum states. In fact, the fundamental physical principles that we explored theoretically encourage the existance of these states in the kind of systems we considered. The main question is if the microscopic parameters of the available materials favor the existance of these states in devices that we can make, or conspire against them. This is a microscopic question that can be ultimately answered only by systematic experiments, or perhaps by ab-initio modeling of electronic spectra and their coupling to phonons in heterostructures. Nevertheless, there are certain phenomenological conclusions that we could make so far. The spin-orbit coupling in topological insulator quantum wells produces a fairly large SU(2) cyclotron energy scale\cite{Nikolic2011a}, which can be of the order of $100 \textrm{ meV}$. A crude calculation in this paper shows that the proximity-induced phonon-mediated interaction in the quantum well alone could be large enough for the two-dimensional pairing to occur in certain heterostructures. These quantum wells are naturally insulating, so if they can be tuned through a phase transition by a gate voltage, their density of mobile spinful particles, triplet Cooper pairs, can be made small and comparable to the SU(2) spin-orbit flux-quantum density\cite{Nikolic2011a}. The quantum fluctuations associated with Cooper pairs are large in this regime, so most essential ingredients for fractional incompressible liquids are present.

What we cannot predict is whether the quantum wells can be made clean enough to obtain sufficiently high mobility (important for fractionalization) and spatially flat potential landscape (important for the gate voltage tunability). We also cannot predict exactly how large the net proximity-induced pairing interactions are in the quantum well. They have to be large enough in order to ensure pairing. The required electron-phonon coupling across the interface can be maximized by eliminating the tunneling barrier in the Fig.\ref{SCTI} and keeping the quantum well thickness $d$ very small. However, this simultaneously diminishes ones ability to control the chemical potential in the quantum well via the gate voltage. It might be necessary to find an optimal barrier and quantum well thickness that achieve a compromise between the electron-phonon coupling and tunability. The choice of materials might also be important, since their interface work functions can make it harder or easier to place the chemical potential in the quantum well near its conduction (or valence) band where bound-state Cooper pair modes should exist.

There are many other kinds of heterostructures and interfaces that we did not address in this paper. One obvious question is whether the proximity effect created by a high-temperature superconductor could give correlated states of matter a better chance. Certainly the conventional proximity effect caused by the U(1) symmetry breaking would be stronger. However, the induced interactions need not be stronger at all. We have found, in fact, a certain inverse relationship between the superconductor's order parameter magnitude (critical temperature) and the strength of proximity-induced interactions, because it is the quantum fluctuations that generate the effective interactions. Furthermore, the cuprate superconductors have a pairing glue whose not-yet-understood dynamics is visible only at very short length-scales and hardly transferable to the other material's surface electrons. In the simple picture of spin-exchange-mediated pairing in cuprates, there is no mechanism for the material's surface electrons to experience the superconductor's spin exchange across the interface (the material's surface is not close to being a half-field fermionic Mott insulator). This eliminates the possibility of the proximity-induced triplet pairing, additionally aggravated by the strongly spin-dependent nature of interactions in cuprates, which is passed on to the material's electrons by tunneling.

Other important proximity effects are created by magnetic rather than superconducting materials. Here, too, we find no significant analogue of the phonon mechanism. Since magnetic orders are stabilized by Coulomb interactions, photons would be the proper substitute for phonons in magnetic proximity effects. The ensuing magnetic interactions are dipole-dipole, typically not very important at short length-scales. The effective surface interactions mediated by electron tunneling can be deduced using the similar approach as in the second part of this paper, but done in the particle-hole channels. We expect a similar phenomenological outcome, that the magnetic material's repulsive electron-electron interactions are imprinted in a weaker form on the surface electrons across the interface. There is, of course, an induced Zeeman coupling as well, the result of symmetry breaking usually considered as the proximity effect. The induced interactions could in principle lead to new surface phases, especially if the surface dynamics creates opportunities for instabilities in the presence of interactions that do not exist in the magnetic material.

\section{Acknowledgements}

The support for this research was provided by the Office of Naval Research (grant N00014-09-1-1025A), and the National Institute of Standards and Technology (grant 70NANB7H6138, Am 001). Work at the Johns Hopkins-Princeton Institute for Quantum Matter was supported by the U.\ S.\ Department of Energy, Office of Basic Energy Sciences, Division of Materials Sciences and Engineering, under Award No.\ DE-FG02-08ER46544.

\appendix
\bigskip

\section{Bethe-Salpeter equation}\label{app}

Here we review the properties of the two-particle Green's function $\mathcal{G}$ that appears in the Bethe-Salpeter equation (\ref{BS1}). Our specific goal is to establish the relationships (\ref{Gcoh}), (\ref{WF2}) and (\ref{BSsol}) that led us to the formulation of the Schrodinger equation for the Cooper pair bound-states. Note that we use the symbol $\tau$ (lacking a better one) to label a time variable in this appendix; this should not be confused with the orbital labels from the rest of the paper.

The two-body Green's function is:
\begin{eqnarray}
&& \!\!\!\!\!\!\! \mathcal{G}(X_{\mu}^{\phantom{k}},x_{\mu}^{\phantom{k}};X_{\mu}',x_{\mu}')=
  -i\langle0|T \psi\!\left(X_{\mu}^{\phantom{k}}+\frac{x_{\mu}^{\phantom{k}}}{2}\right)
    \!\psi\!\left(X_{\mu}^{\phantom{k}}-\frac{x_{\mu}^{\phantom{k}}}{2}\right) \nonumber \\
&& \times \psi^{\dagger}\!\left(X_{\mu}'-\frac{x_{\mu}'}{2}\right)
    \!\psi^{\dagger}\!\left(X_{\mu}'+\frac{x_{\mu}'}{2}\right)|0\rangle \ , \!\!\!\!\!\! \\ \nonumber
\end{eqnarray}
where $\psi^{\dagger},\psi$ are the single-particle creation and annihilation operators in the Heisenberg picture, and $X_{\mu}=({\bf R},\tau)$ are the center-of-mass and $x_{\mu}=({\bf r},t)$ the relative coordinates of the two particles. After expanding the time-ordering, partially applying the time evolution $\psi(\tau\pm t/2)=e^{iH\tau}\psi(\pm t/2)e^{-iH\tau}$ and inserting the identity operator resolved in terms of all exact eigenstates $|n\rangle$ of the many-body Hamiltonian, we obtain:
\begin{widetext}
\begin{equation}
\mathcal{G}(X_{\mu}^{\phantom{k}},x_{\mu}^{\phantom{k}};X_{\mu}',x_{\mu}') = -i\left\lbrace
  \begin{array}{ccc}
    \sum_{p}e^{-iE_{p}|\tau-\tau'|}\mathcal{F}_{p}^{\phantom{\dagger}}({\bf R},{\bf r},t)
      \mathcal{F}_{p}^{\dagger}({\bf R}',{\bf r}',t')
      & , & \tau-\frac{|t|}{2}>\tau'+\frac{|t'|}{2}\\
    \sum_{h}e^{-iE_{h}|\tau-\tau'|}\mathcal{F}_{h}^{\phantom{\dagger}}({\bf R},{\bf r},t)
      \mathcal{F}_{h}^{\dagger}({\bf R}',{\bf r}',t')
      & , & \tau+\frac{|t|}{2}<\tau'-\frac{|t'|}{2}\end{array}\right\rbrace +\cdots \,
\end{equation}
\end{widetext}
where we emphasized only the contributions of two-particle $|p\rangle$ and two-hole $|h\rangle$ excitations, and defined
\begin{eqnarray}
\mathcal{F}_{p}({\bf R},{\bf r},t) \!\!&=&\!\! \langle0|T\psi\left({\bf R}+\frac{{\bf r}}{2},\frac{t}{2}\right)
  \psi\left({\bf R}-\frac{{\bf r}}{2},-\frac{t}{2}\right)|p\rangle \nonumber \\
\mathcal{F}_{h}({\bf R},{\bf r},t) \!\!&=&\!\! \langle h|T\psi\left({\bf R}+\frac{{\bf r}}{2},\frac{t}{2}\right)
  \psi\left({\bf R}-\frac{{\bf r}}{2},-\frac{t}{2}\right)|0\rangle \ . \nonumber
\end{eqnarray}
$E_p$ and $E_h$ are the exact energies of the two-particle and two-hole eigenstates. Taking the Fourier transform over $\tau$ yields:
\begin{eqnarray}
&& \!\!\!\!\!\!\!\! \mathcal{G}({\bf R},\Omega,{\bf r},t;{\bf R}',\Omega',{\bf r}',t') = (2\pi)\delta(\Omega-\Omega')\times\\[0.1in]
&& \!\!\!\!\! \Biggl\lbrack \sum_{p}\frac{\mathcal{F}_{p}^{\phantom{\dagger}}({\bf R},{\bf r},t)
   \mathcal{F}_{p}^{\dagger}({\bf R}',{\bf r}',t')}{\Omega-E_{p}+i0^{+}}e^{i(\Omega-E_{p}+i0^{+})\frac{|t|+|t'|}{2}} - \nonumber \\
&& \!\!\!\!\!  \sum_{h}\frac{\mathcal{F}_{h}^{\phantom{\dagger}}({\bf R},{\bf r},t)
   \mathcal{F}_{h}^{\dagger}({\bf R}',{\bf r}',t')}{\Omega+E_{h}-i0^{+}}e^{-i(\Omega+E_{h}-i0^{+})\frac{|t|+|t'|}{2}}
   +\cdots \Biggr\rbrack \nonumber
\end{eqnarray}
We will be interested in a particular two-particle bound-state $|p\rangle$ with energy $E_{p}=\mathcal{E}$, and set accordingly $\Omega$ arbitrarily close (or at) $\mathcal{E}$:
\begin{eqnarray}
&& \mathcal{G}({\bf R},\Omega,{\bf r},t;{\bf R}',\Omega',{\bf r}',t')\approx
  \frac{\mathcal{F}({\bf R},{\bf r},t)\mathcal{F}^{\dagger}({\bf R}',{\bf r}',t')}{\Omega-\mathcal{E}+i0^{+}} \nonumber \\
&& ~~~~~~~~~~~~~~~~~~~~~~~~~~ \times (2\pi)\delta(\Omega-\Omega') \ .
\end{eqnarray}
The correction to this expression is what we called the incoherent part in (\ref{Gcoh}). Note that the explicit exponential factor involving $|t|+|t'|$ is absent in the written coherent part. We may next carry out the Fourier transform with respect to $t,t'$:
\begin{eqnarray}
&& \mathcal{G}({\bf R},\Omega,{\bf r},\omega;{\bf R}',\Omega',{\bf r}',\omega')\approx
  \frac{\mathcal{F}({\bf R},{\bf r},\omega)\mathcal{F}^{\dagger}({\bf R}',{\bf r}',\omega')}{\Omega-\mathcal{E}+i0^{+}} \nonumber \\
&& ~~~~~~~~~~~~~~~~~~~~~~~~~~ \times (2\pi)\delta(\Omega-\Omega') \ ,
\end{eqnarray}
where
\begin{eqnarray}
&& \mathcal{F}({\bf R},{\bf r},\omega) = \int \dd t\, e^{i\omega t} \times \\
&& ~~~ \langle0|T\psi\left({\bf R}+\frac{{\bf r}}{2},\frac{t}{2}\right)
       \psi\left({\bf R}-\frac{{\bf r}}{2},-\frac{t}{2}\right)|p\rangle \ . \nonumber
\end{eqnarray}
Note that
\begin{eqnarray}
&& \int\frac{\dd\omega}{2\pi}\,\mathcal{F}({\bf R},{\bf r},\omega)=\mathcal{F}({\bf R},{\bf r},t=0) \\
&& ~~~ =\langle0|\psi\left({\bf R}+\frac{{\bf r}}{2}\right)\psi\left({\bf R}-\frac{{\bf r}}{2}\right)|p\rangle
       \equiv \Psi({\bf R},{\bf r}) \nonumber
\end{eqnarray}
is the standard stationary two-body wavefunction of the bound-state with the center-of-mass at ${\bf R}$ and the relative displacement between particles ${\bf r}$. This leads to the equation (\ref{WF2}). In the usual circumstances, we can separate variables $\Psi({\bf R},{\bf r}) = \Psi_{\textrm{cm}}({\bf R}) \psi({\bf r})$, where $\psi({\bf r})$ is the wavefunction of the relative motion of the two particles. Lastly, we may carry out the Fourier transform to the momentum space:
\begin{equation}\label{Gpw}
\mathcal{G}(P_{\mu}^{\phantom{k}},p_{\mu}^{\phantom{k}};P_{\mu}',p_{\mu}')\approx
   \frac{\mathcal{F}({\bf P},p_{\mu}^{\phantom{k}})\mathcal{F}^{\dagger}({\bf P}',p_{\mu}')}{\Omega-\mathcal{E}+i0^{+}}
   (2\pi)^3\delta(P_{\mu}^{\phantom{k}}-P_{\mu}') \ ,
\end{equation}
where
\begin{equation}
\mathcal{F}({\bf P},{\bf p},\omega)=\int \dd^{2}R\, \dd^{2}r\, e^{-i({\bf PR}+{\bf pr})}\mathcal{F}({\bf R},{\bf r},\omega) .
\end{equation}
Note that the total energy $\mathcal{E}$ depends on the total conserved momentum ${\bf P}$ conjugate to the center-of-mass position ${\bf R}$.

The Bethe-Salpeter equation (\ref{BS1}) is obtained by substituting (\ref{Gpw}) into the Dyson equation (\ref{Dyson}), taking the $\Omega \to \mathcal{E}$ limit and neglecting the incoherent parts of the Green's function. Specifically, the bare Green's function $\mathcal{G}_0$ that appears as a separate additive term in the Dyson equation is negligible at the bound-state pole of the exact Green's function $\mathcal{G}$. After this term is neglected, the common factors of $\mathcal{F}^\dagger$ from the exact Green's function can be canceled out since nothing operates on them. Finally, by the separation of variables $\mathcal{F}({\bf P},{\bf p},\omega) = \mathcal{F}_{\textrm{cm}}({\bf P}) F({\bf p},\omega)$ into the center-of-mass and relative coordinates, we can cancel out the common factors of $\mathcal{F}_{\textrm{cm}}({\bf P})$ to obtain (\ref{BS1}).

The function $\mathcal{F}({\bf P},p_{\mu})$ is still a Green's function with a frequency dependence. We can try to elucidate the meaning of this frequency by the same approach that we took to reveal the meaning of the center-of-mass frequency $\Omega$. To that end, we will carry out the time-ordering a bit more carefully than above (assuming fermionic particles):
\begin{widetext}
\begin{eqnarray}
\mathcal{F}({\bf P},p_{\mu}) \!\!&=&\!\!
  \sum_{m}\left\lbrace \begin{array}{ccc}
    \langle0|e^{iHt/2}\psi\left({\bf R}+\frac{{\bf r}}{2}\right)e^{-iHt/2}
      |m\rangle\langle m|
    e^{-iHt/2}\psi\left({\bf R}-\frac{{\bf r}}{2}\right)e^{iHt/2}|p\rangle
  & , & t>0\\
   -\langle0|e^{-iHt/2}\psi\left({\bf R}-\frac{{\bf r}}{2}\right)e^{iHt/2}
      |m\rangle\langle m|
    e^{iHt/2}\psi\left({\bf R}+\frac{{\bf r}}{2}\right)e^{-iHt/2}|p\rangle
  & , & t<0\end{array}\right\rbrace \nonumber \\
 \!\!&=&\!\!
  \sum_{m}e^{i\left(\frac{E_{p}}{2}-E_{m}\right)|t|}\left\lbrace \begin{array}{ccc}
    \langle0|\psi\left({\bf R}+\frac{{\bf r}}{2}\right)
      |m\rangle\langle m|
    \psi\left({\bf R}-\frac{{\bf r}}{2}\right)|p\rangle
  & , & t>0\\
   -\langle0|\psi\left({\bf R}-\frac{{\bf r}}{2}\right)
      |m\rangle\langle m|\psi\left({\bf R}+\frac{{\bf r}}{2}\right)|p\rangle
  & , & t<0\end{array}\right\rbrace  \ .
\end{eqnarray}
\end{widetext}
The sums run over all exact single-particle excitations $|m\rangle$. The Fourier transform yields:
\begin{eqnarray}
&& \mathcal{F}({\bf R},{\bf r},\omega)=i\sum_{m}\Biggl\lbrack
    \frac{\langle0|\psi\left({\bf R}+\frac{{\bf r}}{2}\right)|m\rangle\langle m|\psi\left({\bf R}-\frac{{\bf r}}{2}\right)|p\rangle}
      {\omega-\left(\frac{E_{p}}{2}-E_{m}\right)+i0^{+}} \nonumber \\
&& ~~~~~~
   +\frac{\langle0|\psi\left({\bf R}-\frac{{\bf r}}{2}\right)|m\rangle\langle m|\psi\left({\bf R}+\frac{{\bf r}}{2}\right)|p\rangle}
      {\omega+\left(\frac{E_{p}}{2}-E_{m}\right)-i0^{+}}\Biggr\rbrack \ .
\end{eqnarray}
If the two-body spectrum contains a bound state $|p\rangle$, then all single-particle states $|m\rangle$ have energies $E_{m}$ that lie above the bound-state energy $E_{p}$ by a finite amount, $E_{p}-2E_{m}<0$. The largest $E_{p}-2E_{m}$ is the binding energy $\epsilon<0$, and setting $\omega=\pm\epsilon/2$ should generate a pole. In the absence of a bound-state, the would-be pole is embedded in the two-particle continuum and smeared out by the integral (sum) over $|m\rangle$. In a conventional bulk BCS superconductor, however, the pole is resurrected despite the absence of a Cooper pair bound-state due to the pairing gap $\Delta_{0}$: $E_{p}=0, E_{m}=\sqrt{E^{2}({\bf p})+\Delta_{0}^{2}}$, implying $E_{p}/2-E_{m}<0$ for all $|m\rangle$. Clearly, the pairing gap and a Cooper pair binding energy (when applicable) play the same physical role. The function $F({\bf p},\omega)$ turns into the anomalous propagator in the BCS theory.


\begin{thebibliography}{10}

\bibitem{Read2000}
N. Read and D. Green, Physical Review B {\bf 61},  10267  (2000).

\bibitem{Kitaev2000}
A. Kitaev, Physics-Uspekhi {\bf 44 Supplement},  131  (2000).

\bibitem{Fu2008}
L. Fu and C.~L. Kane, Physical Review Letters {\bf 100},  096407  (2008).

\bibitem{Sato2009}
M. Sato, Y. Takahashi, and S. Fujimoto, Physical Review Letters {\bf 103},
  020401  (2009).

\bibitem{Stanescu2010a}
T.~D. Stanescu, J.~D. Sau, R.~M. Lutchyn, and S.~D. Sarma, Physical Review B
  {\bf 81},  241310(R)  (2010).

\bibitem{Sau2010c}
J.~D. Sau, R.~M. Lutchyn, S. Tewari, and S.~D. Sarma, Physical Review B {\bf
  82},  094522  (2010).

\bibitem{Linder2010}
J. Linder, Y. Tanaka, T. Yokoyama, A. Sudb\o{}, and N. Nagaosa, Physical Review
  Letters {\bf 104},  067001  (2010).

\bibitem{Sato2010}
M. Sato, Y. Takahashi, and S. Fujimoto, Physical Review B {\bf 82},  134521
  (2010).

\bibitem{Sau2010}
J.~D. Sau, R.~M. Lutchyn, S. Tewari, and S.~D. Sarma, Physical Review Letters
  {\bf 104},  040502  (2010).

\bibitem{Sau2010a}
J.~D. Sau, S. Tewari, R.~M. Lutchyn, T.~D. Stanescu, and S.~D. Sarma, Physical
  Review B {\bf 82},  214509  (2010).

\bibitem{Lutchyn2010}
R.~M. Lutchyn, J.~D. Sau, and S.~D. Sarma, Physical Review Letters {\bf 105},
  077001  (2010).

\bibitem{Oreg2010}
Y. Oreg, G. Refael, and F. von Oppen, Physical Review Letters {\bf 105},
  177002  (2010).

\bibitem{Stanescu2011}
T. Stanescu, R.~M. Lutchyn, and S.~D. Sarma, Physical Review B {\bf 84},
  144522  (2011).

\bibitem{Cook2011}
A. Cook and M. Franz, Physical Review B {\bf 84},  201105(R)  (2011).

\bibitem{Alicea2011}
J. Alicea, Y. Oreg, G. Refael, F. von Oppen, and M.~P.~A. Fisher, Nature
  Physics {\bf 7},  412  (2011).

\bibitem{Potter2012}
A.~C. Potter and P.~A. Lee, Physical Review B {\bf 85},  094516  (2012).

\bibitem{Lin2012}
C.-H. Lin, J.~D. Sau, and S.~D. Sarma, Physical Review B {\bf 86},  224511
  (2012).

\bibitem{Fidkowski2012}
L. Fidkowski, J. Alicea, N. Lindner, R.~M. Lutchyn, and M.~P.~A. Fisher,
  Physical Review B {\bf 85},  245121  (2012).

\bibitem{Kells2012}
G. Kells, D. Meidan, and P.~W. Brouwer, Physical Review B {\bf 86},  100503(R)
  (2012).

\bibitem{Stanescu2012a}
T.~D. Stanescu, S. Tewari, J.~D. Sau, and S.~D. Sarma, Physical Review Letters
  {\bf 109},  266402  (2012).

\bibitem{Alicea2012}
J. Alicea, Reports on Progress in Physics {\bf 75},  076501  (2012).

\bibitem{Valls2010}
O.~T. Valls, M. Bryan, and I. Zutic, Physical Review B {\bf 82},  134534
  (2010).

\bibitem{Lababidi2011}
M. Lababidi and E. Zhao, Physical Review B {\bf 83},  184511  (2011).

\bibitem{Rol2011}
R. Grein, J. Michelsen, and M. Eschrig,   (2011), arXiv:1111.0445.

\bibitem{Sau2012}
J.~D. Sau, S. Tewari, and S.~D. Sarma, Physical Review B {\bf 85},  064512
  (2012).

\bibitem{Michelsen2012}
J. Michelsen and R. Grein,   (2012), arXiv:1208.1090.

\bibitem{Mourik2012}
V. Mourik, K. Zuo, S.~M. Frolov, S.~R. Plissard, E.~P. A.~M. Bakkers, and L.~P.
  Kouwenhoven, Science {\bf 336},  1003  (2012).

\bibitem{Deng2012}
M.~T. Deng, C.~L. Yu, G.~Y. Huang, M. Larsson, P. Caroff, and H.~Q. Xu, Nano
  Letters {\bf 12},  6414  (2012).

\bibitem{Rokhinson2012}
L.~P. Rokhinson, X. Liu, and J.~K. Furdyna, Nature Physics {\bf 8},  795
  (2012).

\bibitem{Das2012}
A. Das, Y. Ronen, Y. Most, Y. Oreg, M. Heiblum, and H. Shtrikman, Nature
  Physics {\bf 8},  887  (2012).

\bibitem{Kasumov1996}
A.~Y. Kasumov, O.~V. Kononenko, V.~N. Matveev, T.~B. Borsenko, V.~A. Tulin,
  E.~E. Vdovin, and I.~I. Khodos, Physical Review Letters {\bf 77},  3029
  (1996).

\bibitem{Koren2011}
G. Koren, T. Kirzhner, E. Lahoud, K.~B. Chashka, and A. Kanigel, Physical
  Review B {\bf 84},  224521  (2011).

\bibitem{Qu2011}
F. Qu, F. Yang, J. Shen, Y. Ding, J. Chen, Z. Ji, G. Liu, J. Fan, X. Jing, C.
  Yang, and L. Lu, Scientific Reports {\bf 2},  339  (2012).

\bibitem{Sacepe2011}
B. Sacepe, J.~B. Oostinga, J. Li, A. Ubaldini, N.~J.~G. Couto, E. Giannini, and
  A.~F. Morpurgo, Nature Communications {\bf 2},  575  (2011).

\bibitem{Yang2011}
F. Yang, Y. Ding, F. Qu, J. Shen, J. Chen, Z. Wei, Z. Ji, G. Liu, J. Fan, C.
  Yang, T. Xiang, and L. Lu, Physical Review B {\bf 85},  104508  (2011).

\bibitem{Zhang2011}
D. Zhang, J. Wang, A.~M. DaSilva, J.~S. Lee, H.~R. Gutierrez, M.~H.~W. Chan, J.
  Jain, and N. Samarth, Physical Review B {\bf 84},  165120  (2011).

\bibitem{Veldhorst2012}
M. Veldhorst, M. Snelder, M. Hoek, T. Gang, V.~K. Guduru, X.~L. Wang, U.
  Zeitler, W.~G. van~der Wiel, A.~A. Golubov, H. Hilgenkamp, and A. Brinkman,
  Nature Materials {\bf 11},  417  (2012).

\bibitem{Wang2012}
J. Wang, C.-Z. Chang, H. Li, K. He, D. Zhang, M. Singh, X.-C. Ma, N. Samarth,
  M. Xie, Q.-K. Xue, and M.~H.~W. Chan, Physical Review B {\bf 85},  045415
  (2012).

\bibitem{Wang2012a}
M.-X. Wang, C. Liu, J.-P. Xu, F. Yang, L. Miao, M.-Y. Yao, C.~L. Gao, C. Shen2,
  X. Ma, X. Chen, Z.-A. Xu, Y. Liu, S.-C. Zhang, D. Qian, J.-F. Jia, and Q.-K.
  Xue, Science {\bf 336},  52  (2012).

\bibitem{Williams2012}
J.~R. Williams, A.~J. Bestwick, P. Gallagher, S.~S. Hong, Y. Cui, A.~S. Bleich,
  J.~G. Analytis, I.~R. Fisher, and D. Goldhaber-Gordon, Physical Review
  Letters {\bf 109},  056803  (2012).

\bibitem{Yang2012}
F. Yang, F. Qu, J. Shen, Y. Ding, J. Chen, Z. Ji, G. Liu, J. Fan, C. Yang, L.
  Fu, and L. Lu, Physical Review B {\bf 86},  134504  (2012).

\bibitem{Nikolic2011a}
P. Nikolic, T. Duric, and Z. Tesanovic, Physical Review Letters  (2013), (in
  press).

\bibitem{Nikolic2012a}
P. Nikolic and Z. Tesanovic, Physical Review B {\bf 87},  104514  (2013).

\bibitem{Hsieh2009}
D. Hsieh, Y. Xia, D. Qian, L. Wray, J.~H. Dil, F. Meier, J. Osterwalder, L.
  Patthey, J.~G. Checkelsky, N.~P. Ong, A.~V. Fedorov, H. Lin, A. Bansil, D.
  Grauer, Y.~S. Hor, R.~J. Cava, and M.~Z. Hasan, Nature {\bf 460},  1101
  (2009).

\bibitem{Zhang2010}
Y. Zhang, K. He, C.-Z. Chang, C.-L. Song, L.-L. Wang, X. Chen, J.-F. Jia, Z.
  Fang, X. Dai, W.-Y. Shan, S.-Q. Shen, Q. Niu, X.-L. Qi, S.-C. Zhang, X.-C.
  Ma, and Q.-K. Xue, Nature Physics {\bf 6},  584  (2010).

\bibitem{Zhang2009c}
G. Zhang, H. Qin, J. Teng, J. Guo, Q. Guo, X. Dai, Z. Fang, and K. Wu, Applied
  Physics Letters {\bf 95},  053114  (2009).

\bibitem{Kong2010}
D. Kong, W. Dang, J.~J. Cha, H. Li, S. Meister, H. Peng, Z. Liu, and Y. Cui,
  Nano Letters {\bf 10},  2245  (2010).

\bibitem{Hong2010}
S.~S. Hong, W. Kundhikanjana, J.~J. Cha, K. Lai, D. Kong, S. Meister, M.~A.
  Kelly, Z.-X. Shen, and Y. Cui, Nano Letters {\bf 10},  3118  (2010).

\bibitem{Liu2011a}
M. Liu, C.-Z. Chang, Z. Zhang, Y. Zhang, W. Ruan, K. He, L. li~Wang, X. Chen,
  J.-F. Jia, S.-C. Zhang, Q.-K. Xue, X. Ma, and Y. Wang, Physical Review B {\bf
  83},  165440  (2011).

\bibitem{Cho2011a}
S. Cho, N.~P. Butch, J. Paglione, and M.~S. Fuhrer, Nano Letters {\bf 11},
  1925  (2011).

\bibitem{Moore1989}
M.~A. Moore, Physical Review B {\bf 39},  136  (1989).

\bibitem{Tesanovic1994}
Z. Tesanovic, Physica C {\bf 220},  303  (1994).

\bibitem{Sinova2002}
J. Sinova, C.~B. Hanna, and A.~H. MacDonald, Physical Review Letters {\bf 89},
  030403  (2002).

\bibitem{Nikolic2011}
P. Nikolic, Journal of Physics: Condensed Matter {\bf 25},  025602  (2013).

\bibitem{Nikolic2012}
P. Nikolic,   (2012), arXiv:1206.1055.

\bibitem{L1977}
L.~D. Landau and E.~M. Lifshitz, {\em Course of Theoretical Physics: Quantum
  Mechanics}, 3 ed. (Butterworth-Heinemann, Oxford, 1977), Vol.~3.

\bibitem{R1989}
M. Randeria, J.-M. Duan, and L.-Y. Shieh, Physical Review Letters {\bf 62},
  981  (1989).

\bibitem{SR1989}
S. Schmitt-Rink, C.~M. Varma, and A.~E. Ruckenstein, Physical Review Letters
  {\bf 63},  445  (1989).

\bibitem{Nussinov05}
Z. Nussinov and S. Nussinov, Physical Review A {\bf 74},  053622  (2006).

\bibitem{nikolic:144507}
P. Nikolic, Physical Review B {\bf 79},  144507  (2009).

\bibitem{Nikolic2010}
P. Nikolic, Physical Review B {\bf 83},  064523  (2011).

\bibitem{Nikolic2010b}
P. Nikolic and Z. Tesanovic, Physical Review B {\bf 83},  064501  (2011).

\bibitem{Mahan2000}
G.~D. Mahan, {\em Many-Particle Physics}, {\em Physics of Solids and Liquids},
  3rd ed. ed. (Kluwer Academic / Plenum Publishers, New York, 2000).

\bibitem{Abrikosov1975}
A.~A. Abrikosov, L.~P. Gorkov, and I.~E. Dzyaloshinski,  in {\em Methods of
  Quantum Field Theory in Statistical Physics}, edited by R.~A. Silverman
  (Dover Publications, Inc., New York, 1975).

\bibitem{Ashcroft1976}
N.~W. Ashcroft and N.~D. Mermin, {\em Solid State Physics} (Harcourt College
  Publishers, Orlando, 1976).

\bibitem{Putti2003}
M. Putti, V. Braccini, E. Galleani, F. Napoli, I. Pallecchi, A.~S. Siri, P.
  Manfrinetti, and A. Palenzona, Superconductor Science and Technology {\bf
  16},  188  (2003).

\bibitem{Cappelluti2004}
E. Cappelluti, G.~B. Bachelet, L. Boeri, S. Ciuchi, C. Grimaldi, L. Pietronero,
  and S. Str\"{a}ssler, Physica C {\bf 408},  332  (2004).

\bibitem{Pallecchi2006}
I. Pallecchi, M. Monni, C. Ferdeghini, V. Ferrando, M. Putti, C. Tarantini, and
  E.~G. D'Agliano, European Physical Journal B {\bf 52},  171  (2006).

\end{thebibliography}



\end{document}